\documentclass{article}
\usepackage{PRIMEarxiv}
\usepackage[utf8]{inputenc} 
\usepackage[T1]{fontenc}    
\usepackage{hyperref}       
\usepackage{url}            
\usepackage{booktabs}       
\usepackage{amsfonts}       
\usepackage{nicefrac}       
\usepackage{microtype}      
\usepackage{lipsum}
\usepackage{fancyhdr}       
\usepackage{graphicx}       
\usepackage{xcolor}
\usepackage{bm}
\usepackage{amsmath}
\usepackage{amssymb}
\usepackage{pifont}

\newcommand{\xmark}{\ding{55}}
\usepackage{siunitx}
\usepackage{blindtext}
\usepackage{multirow}
\usepackage{array} 
\usepackage{pdflscape}
\usepackage{booktabs}
\graphicspath{{media/}}    
\usepackage[round]{natbib}
\title{Overcoming Standardization: Revealing Hidden Age Patterns of Suicide with Spatiotemporal Models
}

\author{
  Javier Martín-Pozuelo\\
  Department of Statistics and Operations Research\\ 
  University of Valencia \\
  \texttt{martinpo@alumni.uv.es}\\
  \And
  Antonio López-Quílez \\
  Department of Statistics and Operations Research\\ 
  University of Valencia \\
  \texttt{Antonio.Lopez@uv.es}\\
  \And
  Xavier Barber \\
  Joint Research Unit UMH-FISABIO (StatSalut) \\
  Center of Operations Research \\
  Miguel Hernández University \\
  \texttt{xbarber@umh.es} \\
  \And
  Miriam Marco \\
  Department of Social Psychology\\ 
  University of Valencia \\
  \texttt{miriam.marco2@uv.es}\\
}

\begin{document}
\maketitle

\begin{abstract}
Indirect standardization is widely used in disease mapping to control for confounding, but relies on restrictive assumptions that may bias estimates if violated. Using data on suicide-related emergency calls, this study highlights such limitations and proposes age-structured hierarchical Bayesian models as an alternative. These models incorporate space-time, space-age, and time-age interactions, allowing for more accurate estimation without strong assumptions. The results show improved model fit, especially when including age effects. The best model reveals a rising temporal trend (2017--2022), a nonlinear age pattern, and stronger risk increases among younger individuals compared to older ones.
\end{abstract}


\section{Introduction}

In epidemiology, identifying factors that influence disease patterns-such as age-is critical for accurate risk assessment. For example, many diseases or causes of interest are intrinsically related to age, such as an increased risk of heart disease or the overall incidence of cancer. This influence of age highlights the importance of considering it as a key factor in the onset and progression of various diseases, which is fundamental in epidemiological and public health studies. Traditionally, standardization techniques are used in spatial mortality or incidence studies to control the influence of factors such as age on analytical results \citep{age-standardization_1,age-standardization_2,age-standardization_4}. However, indirect standardization assumes, often unrealistically, that age-specific risks are homogeneous across regions and time periods \citep{Wakefield_1,Wakefield_2}. Violations of this assumption can bias estimates, particularly in space-time analyses. Furthermore, methodological advances in recent decades allow for the incorporation of new effects by integrating different structures, which benefits the application of more detailed studies of the potential and differential impact of these confounding factors, generating higher-quality and more useful statistical results.\\

Rencently, greater accessibility to geographical data has driven a growing interest in spatial statistics. Modern information systems collect increasingly large volumes of data, which increases the need for appropriate statistical methods to obtain valuable information. In parallel, advances in spatial data analysis techniques have facilitated the adoption of more complex models, allowing more sophisticated spatial relationships to be integrated. This progress has strengthened the popularity of spatio-temporal models, among which the proposal of \cite{Knorr-Held} remains one of the most influential at present \citep{Garazi-Retegui,Gayosa, Urdangarin}. However, the development of more complex spatio-temporal models leads to an increase in the number of estimates, which represents a significant challenge. Nevertheless, advances in Bayesian methods, such as the Integrated Nested Laplace Approximation (INLA) \citep{INLA}, have provided innovative solutions that optimise computation times, making the implementation of more sophisticated statistical models feasible \citep{INLA+,BigDM,inlabru}.\\

Several authors have proposed alternatives to indirect standardization with methodologies that seek to incorporate the effect of age in health studies to obtain relevant information. \cite{Congdon} developed an alternative to the proportionality assumption using non-proportional models, obtaining more parsimonious and appropriate models for the area-age dependence. \cite{Standardized_M_Model} use the M-models developed by \cite{M-model} to include the dependence between age and space by means of a correlation matrix. This approach allows reflecting effects in non-separable dependence structures. However, none of these alternatives addresses the possible dependence relationship between time and age. \cite{Age-space-time_CAR} employ models with interactions between their fully structured random effects and they emphasize that this alternative overcomes the assumption that age groups are equally affected by standardization. However, this approach assumes a fully determined structure for the interactions, which may limit its flexibility in contexts where the patterns between space, time, and age do not follow rigid or fully shared structures.\\

In this context, our proposal focuses on a comparative evaluation of the classical approach versus models that integrate age through random effects, with the aim of promoting the use of this modeling as an alternative to standardization. To this end, we will examine in detail the mathematical assumptions that must be met when performing indirect standardization and compare the results with proposals that do not require verifying the proportionality assumption, avoiding erroneous or unreliable risk estimates. Finally, we explain why this methodology constitutes a robust alternative to standardization, even when the proportionality assumption is not violated. To evaluate its applicability in real-life contexts, we apply the proposed methodology to a practical case involving data from calls to the emergency system of the Valencian Community (Spain) related to suicide. Suicide represents a major social and public health challenge that is receiving increasing attention worldwide. Furthermore, several studies have shown that suicide-related emergency calls are not randomly distributed across cities \citep{Suicide_1,Suicide_3,Suicide_4} and that the reasons that lead people to take their own lives can be closely related to age \citep{Suicide_1,Suicide_2}. Therefore, the proposed methodology could reveal hidden patterns that depend on space, age, and time.\\

The article is organized as follows: Section \ref{standardization_procedure} analyzes the proportionality assumption, evaluating its suitability to the case study data. Section \ref{Age_structured models} describes the proposed modeling as an alternative to indirect standardization. Section \ref{Suicide-Related Emergency-Calls Data} presents the results obtained from the analyses performed, and finally, Section \ref{Conclusion} presents the main conclusions of the study.

\section{Indirect Standardization and Risk Proportionality}
\label{standardization_procedure}

Indirect standardization is a classic approach in small-area spatial analysis. It is designed to control for confounding factors by stratifying the population. Despite its historical prevalence and widespread use in statistical modeling, this method relies on restrictive assumptions that often fail in practice, potentially yielding biased or misleading results. This technique is used to incorporate the demographic structure of the regions, taking into account the different strata of the population $k = 1,\dots,K$, through the calculation of expected values.
\begin{equation*}
    E_{ij} = \sum_{k = 1}^{K}N_{ijk} \cdot q_k,
\end{equation*}
where $N_{ijk}$ represents the population at risk in area $i$, period $j$ and stratum $k$, and $q_k$ represents the common incidence in stratum $k$. This formulation assumes a statistical model where the observed counts $O_{ijk}$ follow:
\begin{equation*}
E(O_{ijk}|p_{ijk}) = N_{ijk} \cdot p_{ijk},
\end{equation*}
with $p_{ijk}$ as the stratum-specific incidence probability. In small-area studies, the estimation of $S \times T \times K$ probabilities can be computationally expensive, especially those that involve fine stratification over space and time. This is particularly problematic when the event of interest is rare, as data sparsity can limit the reliability of direct estimation. In such cases, it is often reasonable to assume that the probability of a given unit can be expressed as the product of a spatiotemporal relative risk $\theta_{ij}$ and a base risk $q_k$ associated with the reference stratum, such that.
\begin{equation}\label{decomposition}
    p_{ijk} = \theta_{ij} \cdot q_k.
\end{equation}
Now, within statistical modeling, the Poisson distribution fits perfectly with desirable characteristics, hence a logical assumption is that
\begin{equation*}
O_{ijk}|p_{ijk} \sim Poisson(N_{ijk} \cdot p_{ijk}),
\end{equation*}
where applying the previous decomposition (\ref{decomposition}) and collapsing the units we can arrive at
\begin{equation*}
O_{ij}|\theta_{ij} \sim Poisson(\theta_{ij} \cdot E_{ij}),
\end{equation*}
with $O_{ij} = \sum_{k = 1}^{K}O_{ijk}$ and $E_{ij} = \sum_{k = 1}^{K}{N_{ijk}\cdot q_k}$. All this mathematical development implies that indirect standardization implicitly assumes that 
\begin{equation}
\label{Eq:proportional_assumption_equation}
E \left( \frac{O_{ijk}}{N_{ijk}} \right) = \theta_{ij} \cdot q_{k}.
\end{equation}
Thus, the effect of being in area $i$ at time $j$ is to multiply each of the specific reference rates for each stratum $q_j$ by the relative risk in area-time $\theta_{ij}$. Failure to comply with this assumption can lead to an incorrect estimate of the summarized relative risks \citep{Wakefield_2}. To verify this hypothesis, \cite{Wakefield_1} presents an exploratory test that consists of graphically representing the observed rates by regional stratum $O_{ijk}/N_{ijk}$ against the risk of stratum $q_j$, which allows evaluating a linear relationship between both.

\section{Age-structured models}
\label{Age_structured models}

In this section, we present the age-structured hierarchical Bayesian spatiotemporal models as an alternative to age-standardized models. We assume a Poisson distribution to model the call rate using a hierarchical Bayesian framework. The general structure of the model is described below, followed by some technical specifications such as the necessary restrictions to ensure the identifiability of the fitted models. We consider a division into $S$ geographical units, indexed as $i = 1, \dots, S$, $T$ time periods, indexed as $j = 1, \dots, T$, and $K$ age groups, indexed as $k = 1, \dots, K$. Let $O_{ijk}$ be the observed suicide-related emergency calls, $E_{ijk}$ be the expected values, and $\Theta_{ijk}$ be the relative risk for geographic unit $i$, time period $j$, and age group $k$. We assume that the number of calls to the emergency service follows a Poisson distribution.
\begin{equation}  
\begin{split}
O_{ijk}|\Theta_{ijk} &\sim Poisson(E_{ijk} \Theta_{ijk}),\\ 
\log(\Theta_{ijk}) &= \alpha + \phi_i + \delta_j + \gamma_k + \Lambda_{ijk},
\end{split}
\label{Eq:Model_likelihood}
\end{equation} 
where $\alpha$ is the intercept that will estimate the overall rate, $\phi_i$, $\delta_j$ and $\gamma_k$ are the individual spatial, temporal and age random effects respectively, and $\Lambda_{ijk}$ represents random interaction effects. We will assume a conditional autoregressive distribution of \cite{LCAR} for the spatial parameter $\bm{\phi} = (\phi_1, \dots, \phi_S)'$, a multivariate normal proposal for the temporal effect $\bm{\delta} = (\delta_1, \dots, \delta_T)'$ and for the age effect $\bm{\gamma} = (\gamma_1, \dots, \gamma_T)'$.
\begin{equation*}
\begin{split}
\bm{\phi} &\sim N(\bm{0}, [\tau_\phi (\lambda_\phi \bm{R}_\phi + (1-\lambda_\phi)\bm{I}_\phi)]^{-1}),\\
    \bm{\delta} &\sim N(\bm{0},\tau_\delta^{-1} \bm{R}_\delta ^{-}),\\
    \bm{\gamma} &\sim N(\bm{0},\tau_\gamma^{-1} \bm{R}_\gamma ^{-}),
\end{split}
\label{Eq:Model_Main_effects}
\end{equation*} 
where $\bm{R}_\phi$ represents the spatial structure matrix corresponding to the undirected graph of the spatial regions and their neighbourhood relations, $\bm{I}_\phi$ is the identity matrix of dimension $S \times S$, $\bm{R}_\delta^{-}$ represent the generalised Moore-Penrose inverse of the structured or unstructured matrix of the time effect $\bm{R}_\delta$ and $\tau_\phi$, $\tau_\delta$ and $\tau_\gamma$ are the precision parameters of the spatial, temporal and age effects. We consider both an unstructured effect $\bm{R} = \bm{I}$ and a structured first-order random walk effect (RW1) \citep{RW1_Str} for time and age effects.
\begin{equation} 
   \bm{R} =  \begin{bmatrix}
       1 & -1  \\
       -1 & 2 & -1  \\
        & -1 & 2 & -1 \\
        & & \ddots & \ddots & \ddots \\
        & & & -1 & 2 & -1 \\
        & & & & -1 & 2 & -1 \\
        & & & & & -1 & 1
    \end{bmatrix}.
    \label{Eq:RW_matrix}
\end{equation}
The matrix (\ref{Eq:RW_matrix}) can be seen as the structure matrix of an ICAR-type distribution for $\bm{\delta}$ and $\bm{\gamma}$, analogous to $\bm{\phi}$ in a \cite{Besag} model. The term $\bm{\Lambda}$ includes in an additive way the random interaction effects between the spatial, temporal and age structures. The inclusion of the main effects ($\bm{\phi}$, $\bm{\delta}$ and $\bm{\gamma}$) allows the trends of each term in the linear predictor of $\bm{\Theta}$ to be analysed. Up to this point, this approach could be used with age-standardized data (removing the main effect of age $\bm{\delta}$), following the procedure described in the Section \ref{standardization_procedure}, which would generate additive models on a logarithmic scale under the assumption (\ref{Eq:proportional_assumption_equation}). This work seeks to overcome this inherent limitation of the standardization method and its implied assumptions by incorporating interaction terms, which in turn will identify space-age and time-age patterns. To do so, it is essential to incorporate the interaction effects of the parameter $\bm{\Lambda}$ into the equation (\ref{Eq:Model_likelihood}), adding structures that link the relationships between space, time and age.

\subsection{Incorporating Age effect through Structured Interactions}

Following the framework proposed in \cite{Knorr-Held}, we incorporate age-related interactions into the full modeling framework. Specifically, we include the space-time interaction $\zeta_{ij}^{1}$, the space-age interaction $\zeta_{ik}^{2}$, and the time-age interaction $\zeta_{jk}^{3}$, each of which is assumed to follow a multivariate normal distribution. This allows for a more flexible representation of joint effects across spatial, temporal, and age dimensions.
\begin{equation*} 
\bm{\zeta} \sim N(\bm{0},\tau_\zeta^{-1} \bm{R}_\zeta^{-}),
\label{Eq:Interaction_Dist}
\end{equation*}
where $\tau_\zeta$ is the precision parameter and $\bm{R}_\zeta$ is the matrix obtained by interweaving the dependence structures of the main effects. The interweaving of these structures is performed by the Kronecker product of the structure matrices of the corresponding effects. Table \ref{Tab:Correlation_Interaction} shows the correlations involved for each of the interaction effects:\\

\begin{table}[h!]
\caption{Space-time, space-age and time-age interaction types and their correlations involved.}
\centering
\begin{tabular}{cllccc}
\hline
Interaction & Type & $\bm{R_\zeta}$ & Spatial correlation & Temporal correlation & Age correlation\\ 
  \hline
\multirow{4}{*}{$\bm{\zeta}^1$} & Type I & $\bm{I}_{\phi} \otimes \bm{I}_{\delta}$ & \xmark & \xmark & - \\ 
                   & Type II & $\bm{I}_{\phi} \otimes \bm{R}_{\delta}$ & \xmark & \checkmark & - \\
                   & Type III & $\bm{R}_{\phi} \otimes \bm{I}_{\delta}$ & \checkmark & \xmark & - \\
                   & Type IV & $\bm{R}_{\phi} \otimes \bm{R}_{\delta}$ & \checkmark & \checkmark & - \\
                   \hline
\multirow{4}{*}{$\bm{\zeta}^2$} & Type I & $\bm{I}_{\phi} \otimes \bm{I}_{\gamma}$ & \xmark & - & \xmark \\
                   & Type II & $\bm{I}_{\phi} \otimes \bm{R}_{\gamma}$ & \xmark & - & \checkmark \\
                   & Type III & $\bm{R}_{\phi} \otimes \bm{I}_{\gamma}$ & \checkmark & - & \xmark \\
                   & Type IV & $\bm{R}_{\phi} \otimes \bm{R}_{\gamma}$ & \checkmark & - & \checkmark \\
                   \hline
\multirow{4}{*}{$\bm{\zeta}^3$} & Type I & $\bm{I}_{\gamma} \otimes \bm{I}_{\delta}$ & - & \xmark & \xmark \\
                   & Type II & $\bm{R}_{\gamma} \otimes \bm{I}_{\delta}$ & - & \xmark & \checkmark \\
                   & Type III & $\bm{I}_{\gamma} \otimes \bm{R}_{\delta}$ & - & \checkmark & \xmark \\
                   & Type IV & $\bm{R}_{\gamma} \otimes \bm{R}_{\delta}$ & - & \checkmark & \checkmark \\
                   \hline
\end{tabular}
\label{Tab:Correlation_Interaction}
\end{table}

As is the case for main effects, the structure of the matrix $\bm{R}_{\zeta}$ determines the form of the dependency relationships in the model. The combination of the effects involved gives rise to the different types of structure summarized in Table \ref{Tab:Correlation_Interaction}. The type I interaction of the parameter $\bm{\zeta}^2$ does not imply spatial or age dependence; therefore, the matrix $\bm{R}_{\zeta^2}$ takes the form of a diagonal matrix of dimensions $SA \times SA$, derived from the product of the precision matrices of the unstructured effects in the spatial and age terms. In contrast, the type IV interaction represents the most complex interaction, assuming an age trend for each geographic area. In this case, the vector $\bm{\zeta}^2 = (\zeta^2_{i1}, \dots, \zeta^2_{iA})'$ is assigned an effect structured as a random walk. With this pattern, independence is lost, as the age evolution in area $i$ depends on the trend in neighboring areas. This behavior combines the structured effects of both terms through the product $\bm{R}_{\phi} \otimes \bm{R}_\gamma$, thus integrating the spatial neighborhood structure with the age pattern. The incorporation of these effects is introduced following the general scheme (\ref{Eq:Model_likelihood}) such that $\Lambda_{ijk} = \zeta^1_{ij} + \zeta^2_{ik} + \zeta^3_{jk}$. However, as shown in \cite{Restricciones_zeta1y2}, such effects must be accompanied by their respective constraints\footnote{For the space-age effects $\bm{\zeta}^2$ and time-age effects $\bm{\zeta}^3$, restrictions equivalent to $\bm{\zeta}^1$ have been included in the models.} to ensure the identifiability of the model parameters. Furthermore, these restrictions are related to the rank deficiency associated with the matrices $\bm{R}$ of the interaction effects, with the exception of type I interactions. Table \ref{Tab:Constrain_Interaction} shows a summary of the different restrictions, as well as the rank deficiency of the precision matrices associated with each interaction to ensure the identifiability of the models.\\
\begin{table}[h!]
\caption{Constraint for the identifiability of LCAR models associated with the interaction effects $\bm{\zeta}^1$, $\bm{\zeta}^2$ and $\bm{\zeta}^3$.}
\centering
\begin{tabular}{cllll}
\hline
Interaction & Type & Rank def. & $\bm{R_\zeta}$ & Constraints\\ 
  \hline
\multirow{8}{*}{$\bm{\zeta}^1$} & \multirow{2}{*}{Type I} & \multirow{2}{*}{-} & \multirow{2}{*}{$\bm{I}_{\phi} \otimes \bm{I}_{\delta}$} & $\sum_{i = 1}^{S}\phi_i = 0, \hspace{0.25cm} \sum_{j = 1}^{T}\delta_j = 0 \hspace{0.25cm}$ \\
                                & & & & $\sum_{i = 1}^{S}\sum_{j 
                                = 1}^{T}\zeta_{ij}^{1} =  0$\\
                                & \multirow{2}{*}{Type II} & \multirow{2}{*}{$S$} & \multirow{2}{*}{$\bm{I}_{\phi} \otimes \bm{R}_{\delta}$} & $\sum_{i = 1}^{S}\phi_i = 0, \hspace{0.25cm} \sum_{j = 1}^{T}\delta_j = 0$ \\
                                & & & & $\sum_{j = 1}^{T}\zeta_{ij}^{1} = 0, \hspace{0.25cm} \forall i \in \{1, \dots, S\}$ \\
                                & \multirow{2}{*}{Type III} & \multirow{2}{*}{$T$} & \multirow{2}{*}{$\bm{R}_{\phi} \otimes \bm{I}_{\delta}$} & $\sum_{i = 1}^{S}\phi_i = 0, \hspace{0.25cm} \sum_{j = 1}^{T}\delta_j = 0 $\\
                                & & & & $\sum_{i = 1}^{S}\zeta_{ij}^{1} = 0, \hspace{0.25cm} \forall j \in \{1, \dots, T\}$\\
                                & \multirow{2}{*}{Type IV} & \multirow{2}{*}{$S+T-1$} & \multirow{2}{*}{$\bm{R}_{\phi} \otimes \bm{R}_{\delta}$} & $\sum_{i = 1}^{S}\phi_i = 0, \hspace{0.25cm} \sum_{j = 1}^{T}\delta_j = 0$\\
                                & & & & $\sum_{j = 1}^{T}\zeta_{ij}^{1} = 0, \hspace{0.25cm} \forall i \in \{1, \dots, S\}, \hspace{0.25cm}
                                    \sum_{i = 1}^{S}\zeta_{ij}^{1} = 0, \hspace{0.25cm} \forall j \in \{1, \dots, T\}$\\
\hline                      
\end{tabular}
\label{Tab:Constrain_Interaction}
\end{table}

We examined all possible structures for modeling both the main effects of time and age and all interactions, which involved fitting 520 models. We considered interactions only when their main effects were present, following a sequential strategy similar to model nesting. For simplicity, we chose to use only the Leroux conditional autoregressive distribution for the spatial effect.
\begin{equation} 
\begin{split}
\log(\Theta_{ijk}) &= \alpha + \phi_i + \delta_j,\\
\log(\Theta_{ijk}) &= \alpha + \phi_i + \delta_j + \zeta_{ij}^1,\\
\log(\Theta_{ijk}) &= \alpha + \phi_i + \gamma_k,\\
\log(\Theta_{ijk}) &= \alpha + \phi_i + \gamma_k + \zeta_{ik}^2,\\
\log(\Theta_{ijk}) &= \alpha + \phi_i + \delta_j + \gamma_{k},\\
\log(\Theta_{ijk}) &= \alpha + \phi_i + \delta_j + \gamma_{k} + \zeta_{ij}^1,\\
\log(\Theta_{ijk}) &= \alpha + \phi_i + \delta_j + \gamma_{k} + \zeta_{ij}^1 + \zeta_{ik}^2,\\
\log(\Theta_{ijk}) &= \alpha + \phi_i + \delta_j + \gamma_{k} + \zeta_{ij}^1 + \zeta_{ik}^2 + \zeta_{jk}^3.\\
\end{split}
\label{Eq:Nested_framework}
\end{equation} 

\subsection{Bayesian inference and model selection with INLA}

All models were estimated within a Bayesian framework using Integrated Nested Laplace Approximation (INLA) \citep{INLA}, with penalized complexity priors (PC priors) \citep{PC_priors} for the hyperparameters. These priors are commonly used in INLA due to their regularization properties. To assess the sensitivity of the model to the choice of priors, we also performed a comparison using non-informative priors, verifying that the resulting posterior distributions were robust and consistent (see Table \ref{Tab:Posterior_summary}).\\

INLA obtains approximations of the posterior distribution using strategies based on Laplace approximations coupled with numerical integration techniques. This methodology allows for obtaining accurate estimates of quantities of interest efficiently and without the need for large simulation runs, unlike other methods such as Markov chain Monte Carlo (MCMC). Thanks to this combination of accuracy and computational speed, INLA is especially useful in the analysis of high-dimensional hierarchical models, such as those applied in spatio-temporal analysis or other contexts where the computational burden may be high.\\

We used the Watanabe-Akaike Information Criterion (WAIC) \citep{WAIC} as the primary criterion to compare fitted models. Although several alternatives exist, such as the Deviance Information Criterion (DIC) \citep{DIC} or the Log-Score (LS) \citep{LS}, we opted for WAIC due to its theoretical advantages and empirical stability. Specifically, we used WAIC instead of DIC because some studies suggest that WAIC is more stable in spatio-temporal studies \citep{WAIC_just_1,WAIC_just_2}.\\

All calculations were performed in R version 4.4.2 (10/31/2024) using the INLA package version 24.12.11 (12/11/2024), which provides efficient and accurate approximate Bayesian inference for latent Gaussian models.

\section{Suicide-Related Emergency-Calls Data}
\label{Suicide-Related Emergency-Calls Data}

Suicide represents a growing global public health challenge and remains one of the leading causes of death worldwide. Recent trends show concerning increases in suicide rates, though with significant variations across different age groups. Available evidence consistently indicates particular vulnerability among youth and young adults, who demonstrate proportionally greater rate increases compared to other age demographics. These patterns highlight the critical need for differentiated prevention strategies tailored to the most affected population groups.\\

Some recent studies have taken a spatio-temporal approach to suicide-related data \citep{Suicide_1,Suicide_2}, also showing differentiated patterns by age group. The dependence of the increase in suicide cases on age highlights the need to use more advanced methodologies to extract useful information for public health decision-making. In this context, the models described in the previous section can provide a more detailed analysis of this phenomenon by capturing age-dependent spatial and temporal patterns. This would make it possible to identify with greater precision the geographic regions with the highest risk of suicide according to age group, thus facilitating the design of more focused and effective prevention strategies.\\

\begin{figure}[h!]
    \centering
    \includegraphics[scale=0.45]{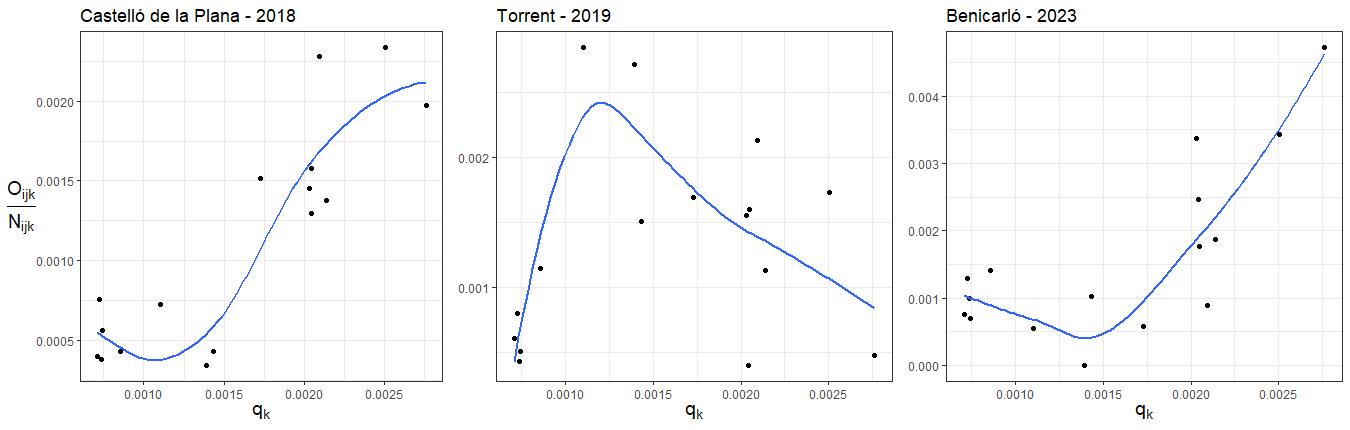}
    \caption{Rates by regional stratum $O_{ijk}/N_{ijk}$ against the risk of stratum $q_j$ of some municipalities-year}
    \label{fig:Proportional_assumption}
\end{figure}

This section presents the results obtained from the space-time-age modeling of suicide-related emergency calls rates in the Valencian Community. First, a graphical analysis is performed to test whether our data support the proportionality of hazards hypothesis described in section \ref{standardization_procedure}. Different municipalities and years were identified in which a violation of the proportionality assumption is evident. Figure \ref{fig:Proportional_assumption} shows a scatterplot of the observed rates by stratum as a function of the stratum's risk, along with a Loess-smoothed curve for some municipality-years where this assumption is not met.

Numerous municipalities have been found where this hypothesis is not verified, indicating how restrictive it can be, especially with data disaggregated in space, time, and age. The property described by equation (\ref{Eq:proportional_assumption_equation}) should be verified for all indices $i$ and $j$. However, the relationship between observed rates and stratum risk clearly deviates from the expected linearity, indicating a possible violation of the assumption. These figures show that according to \cite{Wakefield_1}, the risk estimates obtained using age-standardized models could be erroneous. To obtain more precise estimates of the incidence of calls due to violation of the proportionality assumption, as previously described, different space-time-age conditional autoregressive models are compared to evaluate the effect of geographic, temporal, and age structure, as well as their interactions. To do this, the WAIC results are compared between the best model from each subset of models in the equation (\ref{Eq:Nested_framework}), which allows the impact of incorporating each effect on model fit to be assessed. The estimated main effects and interaction effects are presented below.
\begin{table}[ht!]
\caption{Optimal model for each subset of space-time-age models.}
\centering
\begin{tabular}{rllrrrr}
  \hline
Model & $\bm{\delta}$ & $\bm{\gamma}$ & $\bm{\zeta}^{1}$ & $\bm{\zeta}^{2}$ & $\bm{\zeta}^{3}$ & WAIC\\ 
  \hline
 1 & RW1 & - & - & - & - & 78840.97 \\ 
  2 & RW1 & - & Type IV & - & - & 77854.97 \\
  3 & - & iid & - & - & - & 74362.06 \\ 
  4 & - & RW1 & - & Type II & - & 73411.06 \\
  5 & RW1 & RW1 & - & - & - & 70531.86 \\ 
  6 & RW1 & iid & Type II & - & - & 69225.91 \\
  7 & RW1 & RW1 & Type II & Type II & - & 67680.07 \\ 
  8 & iid & RW1 & Type II & Type I & Type IV & 67505.20 \\ 
   \hline
\end{tabular}
\label{Tab:Best_Models}
\end{table}

Combining all the structures described above generates a total of 520 models (see Supplementary Material \ref{Suplementary_material}). Table \ref{Tab:Best_Models} presents the models with the lowest WAIC among all possible combinations of the effects included in the linear predictor in equation (\ref{Eq:Nested_framework}). Starting from model 1, the incorporation of the space-time effect provides a reduction of 986 units, on the other hand, including only the unstructured effect of age without interaction, as in model 3, already offers an improvement of 3492.91 units compared to model 2. The incorporation of more complex relationships further improves the fit of the models, as in the transition from model 4 to model 5 with an improvement of 2879.20 units, or when incorporating the space-age interaction (Model 7) and the time-age interaction (Model 8) with improvements of 1545.84 and 174.87 units over models 6 and 7 respectively. Furthermore, the incorporation of all effects offers a total improvement of 11,335.77 units, providing strong evidence for the improvement in model fit offered by this methodology. The model with the lowest WAIC, and therefore the one that best fits the data, is Model 8. This model features an unstructured time effect, an age effect with a first-order random walk structure, and combines Type II, I, and IV structures for the space-time, space-age, and time-age interaction effects, respectively. Consequently, it suggests a temporal relationship for the space-time interaction, an unstructured interaction effect between space and age, and a combined effect associated with the time-age interaction.\\
\begin{table}[ht]
\caption{Summary statistics of the posterior distributions of the final space-time-age model.}
\centering
\begin{tabular}{lrrrrrrrr}  
\toprule
\multicolumn{3}{c}{} & \multicolumn{5}{c}{Quantiles} & \multicolumn{1}{c}{} \\
\cmidrule(r){4-8}
parameter & mean & sd & 0.025 & 0.25 & 0.5 & 0.75 & 0.975 & mode \\ 
\midrule
\textbf{Non-Informative priors} &  &  &  &  &  &  &  &  \\ 
$\alpha \hspace{0.40cm} N(\mu = 0, \tau = 0)$ & -0.75 & 0.04 & -0.83 & -0.76 & -0.75 & -0.74 & -0.67 & -0.75 \\
$\sigma_{\phi} \hspace{0.25cm} U(0,\infty)$  & 0.75 & 0.07 & 0.63 & 0.71 & 0.75 & 0.80 & 0.90 & 0.74 \\ 
$\lambda_\phi \hspace{0.25cm} U(0,1)$  & 0.45 & 0.18 & 0.11 & 0.39 & 0.44 & 0.50 & 0.85 & 0.43 \\
$\sigma_{\delta} \hspace{0.25cm} U(0,\infty)$  & 0.37 & 0.09 & 0.23 & 0.31 & 0.36 & 0.43 & 0.59 & 0.33 \\ 
$\sigma_{\gamma} \hspace{0.25cm} U(0,\infty)$ & 0.40 & 0.07 & 0.28 & 0.35 & 0.39 & 0.44 & 0.56 & 0.38 \\ 
$\sigma_{\zeta^1} \hspace{0.12cm} U(0,\infty)$  & 0.25 & 0.01 & 0.23 & 0.25 & 0.25 & 0.26 & 0.28 & 0.25 \\ 
$\sigma_{\zeta^2} \hspace{0.12cm} U(0,\infty)$ & 0.35 & 0.01 & 0.32 & 0.34 & 0.35 & 0.35 & 0.37 & 0.35 \\
$\sigma_{\zeta^3} \hspace{0.12cm} U(0,\infty)$ & 0.05 & 0.01 & 0.04 & 0.05 & 0.05 & 0.06 & 0.08 & 0.05 \\ 
\hline

\textbf{Penalized Complexity priors} &  &  &  &  &  &  \\ 
$\alpha \hspace{0.40cm} N(\mu = 0, \tau = 0)$ & -0.75 & 0.04 & -0.83 & -0.76 & -0.75 & -0.74 & -0.67 & -0.75 \\
$\sigma_{\phi} \hspace{0.25cm} P(\sigma_\phi >1.00) =0.01$  & 0.73 & 0.06 & 0.62 & 0.69 & 0.73 & 0.77 & 0.86 & 0.73 \\ 
$\lambda_\phi \hspace{0.25cm} P(\lambda_\phi < 0.5) =0.5$ & 0.42 & 0.16 & 0.11 & 0.36 & 0.41 & 0.46 & 0.80 & 0.40 \\ 
$\sigma_{\delta} \hspace{0.25cm} P(\sigma_\delta >1.00) =0.01$  & 0.33 & 0.09 & 0.19 & 0.26 & 0.32 & 0.38 & 0.55 & 0.29 \\ 
$\sigma_{\gamma} \hspace{0.25cm} P(\sigma_\gamma >1.00) =0.01$  & 0.38 & 0.07 & 0.27 & 0.33 & 0.37 & 0.42 & 0.53 & 0.36 \\ 
$\sigma_{\zeta^1} \hspace{0.12cm} P(\sigma_{\zeta^1} >1.00) =0.01$  & 0.25 & 0.01 & 0.23 & 0.25 & 0.25 & 0.26 & 0.28 & 0.25 \\ 
$\sigma_{\zeta^2} \hspace{0.12cm} P(\sigma_{\zeta^2} >1.00) =0.01$  & 0.34 & 0.01 & 0.32 & 0.34 & 0.34 & 0.35 & 0.37 & 0.34 \\ 
$\sigma_{\zeta^3} \hspace{0.12cm} P(\sigma_{\zeta^3} >1.00) =0.01$ & 0.05 & 0.01 & 0.04 & 0.05 & 0.05 & 0.06 & 0.08 & 0.05 \\ 
\bottomrule
\end{tabular}
\label{Tab:Posterior_summary}
\end{table}

Regarding the comparison between the posterior distributions according to the selected prior distributions, Table \ref{Tab:Posterior_summary} shows the summary statistics of the posterior distributions obtained by incorporating each type of prior. A slight difference is observed in the summary statistics of the posteriors for some parameters. Specifically, the greatest difference is found in the posterior distribution of the main effects ($\bm{\phi}$, $\bm{\delta}$, and $\bm{\gamma}$), highlighting the standard deviation associated with the time parameter $\bm{\delta}$, with differences in all summary statistics. The posterior distributions associated with the interactions show a much greater consensus. The Leroux distribution parameter $\lambda_\phi$ (Table \ref{Tab:Posterior_summary}) generates a posterior distribution with a mean and a mode concentrated around 0.46 and 0.43, respectively, so no particularly dominant spatial pattern is observed.\\
\begin{figure}[h!]
    \centering
    \includegraphics[width=\linewidth]{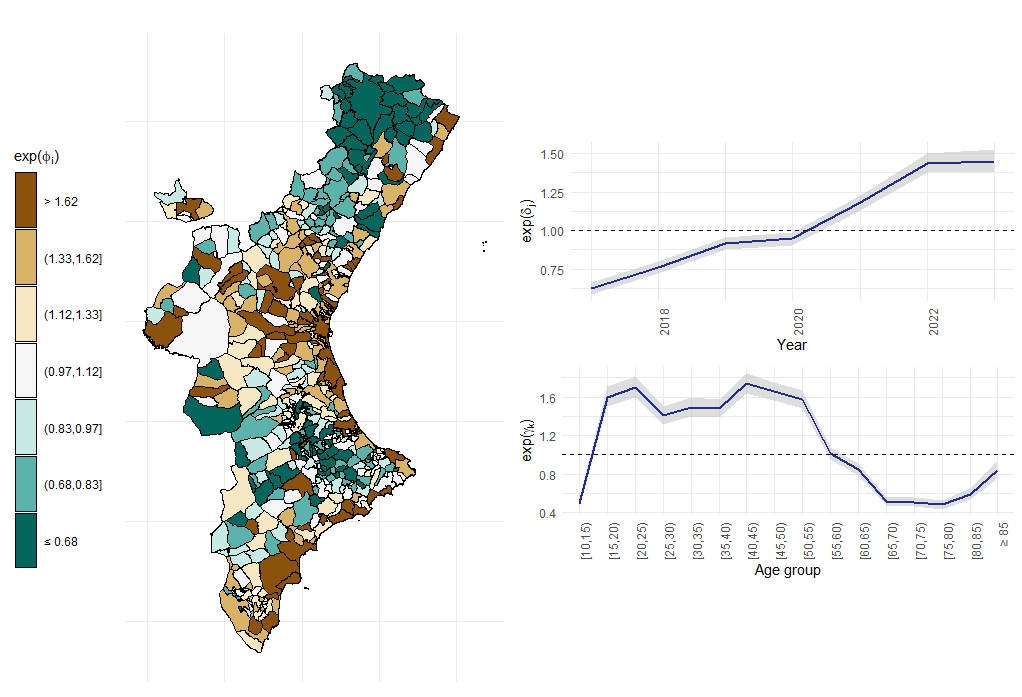}
    \caption{Posterior mean of the main random effects: spatial effect across the Valencian Community ($\exp(\bm{\phi})$), temporal effect over the period 2017--2023 ($\exp(\bm{\delta})$), and age-group effect ($\exp(\bm{\gamma})$).}
    \label{Fig:Main_effects}
\end{figure}

Figure \ref{Fig:Main_effects} shows the main spatial $\exp(\phi_i)$ (left side), temporal $\exp(\delta_j)$ (top right corner), and age $\exp(\gamma_k)$ (bottom right corner) effects. Regarding the spatial effect, the brown regions indicate values greater than 1, implying a positive contribution to the increase in the suicide-related emergency call rate, while the green areas show values less than 1, reflecting a contribution to the decrease in call rate. A large region with a greater contribution is observed in the central area of the map, especially concentrated in the coastal strip (eastern area). Regarding the upper region (Castellón province), the effect is weaker in the inland areas (left side), intensifying along the coast. The lower region (Alicante province) follows a similar pattern, with higher values in the coastal area and lower inland. Regarding the time effect, a practically linear upward trend is identified from 2017 to 2022, followed by a stabilization period between 2022 and 2023. During the 2017--2020 period, the values are less than 1, indicating a contribution to the reduction in the call rate. On the contrary, between 2021 and 2023, the values exceed unity, suggesting an increase in risk. Regarding the age effect, a notable increase is observed in the youngest groups ([15, 20) and [20, 25)), with a positive effect remaining until the group [50, 55). From there, there is a sharp decline in the call rate until the group [75, 80), followed by an upturn in the older age groups. Generally speaking, age groups [15, 55) contribute to the increase in the rate, while groups [10, 15) and those over 55 have a decreasing influence on it.\\
\begin{figure}[ht]
    \centering
    \includegraphics[width=\linewidth]{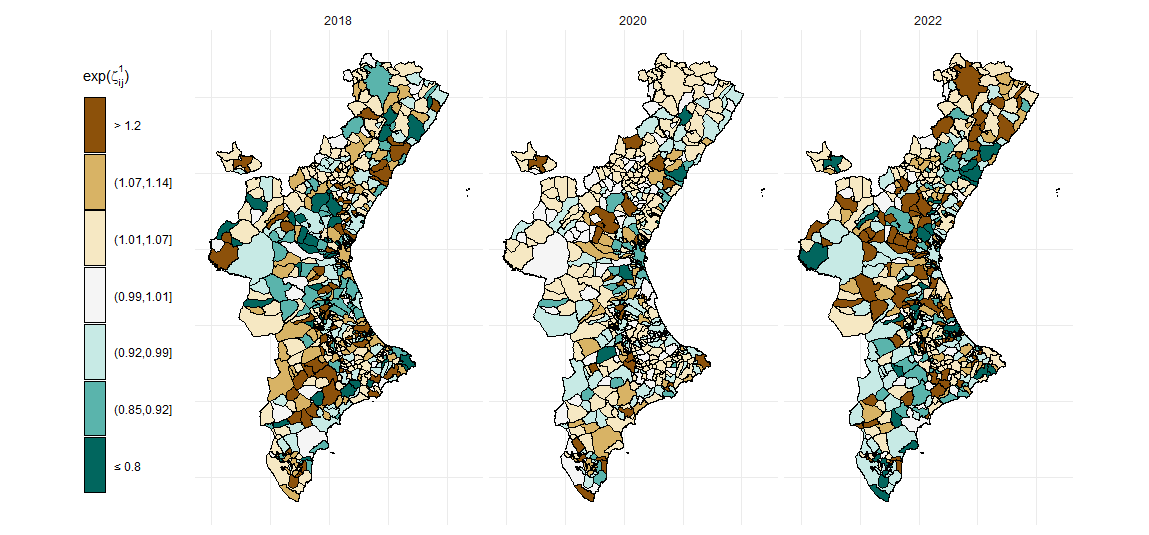}
    \caption{Posterior mean of the spatio-temporal effect ($\exp(\bm{\zeta}^1)$) across the Valencian Community for the years 2018, 2020, and 2022.}
    \label{Fig:Spatio-temporal_effect}
\end{figure}

Figure \ref{Fig:Spatio-temporal_effect} shows the spatio-temporal effect $\exp(\zeta^1_{ij})$ corresponding to the years 2018, 2020, and 2022 (see the complete evolution in the supplementary material). This effect presents a Type II structure, built from an unstructured spatial component and a structured temporal component. This formulation implies that the pattern is not spatially smoothed, but rather reflects heterogeneous spatial variability. Following an interpretation analogous to that of the main spatial effect, the regions in green represent values below 1, indicating a contribution to the reduction in the reduction of the suicide-related emergency calls rate, while the areas in brown show values above 1, implying a contribution to its increase. A progressive decrease in the spatio-temporal effect is observed across a large part of the southern regio of the map (inland Alicante). The central region (Valencia Province) shows a large part of the region with an effect of less than 1 in 2018, as well as a relatively general increase in 2020, and a completely unstructured pattern in 2022. Regarding the upper region (Castellón Province), it is observed that the risk increases especially in the northern area from 2018 to 2022.\\
\begin{figure}[ht]
    \centering
    \includegraphics[width=\linewidth]{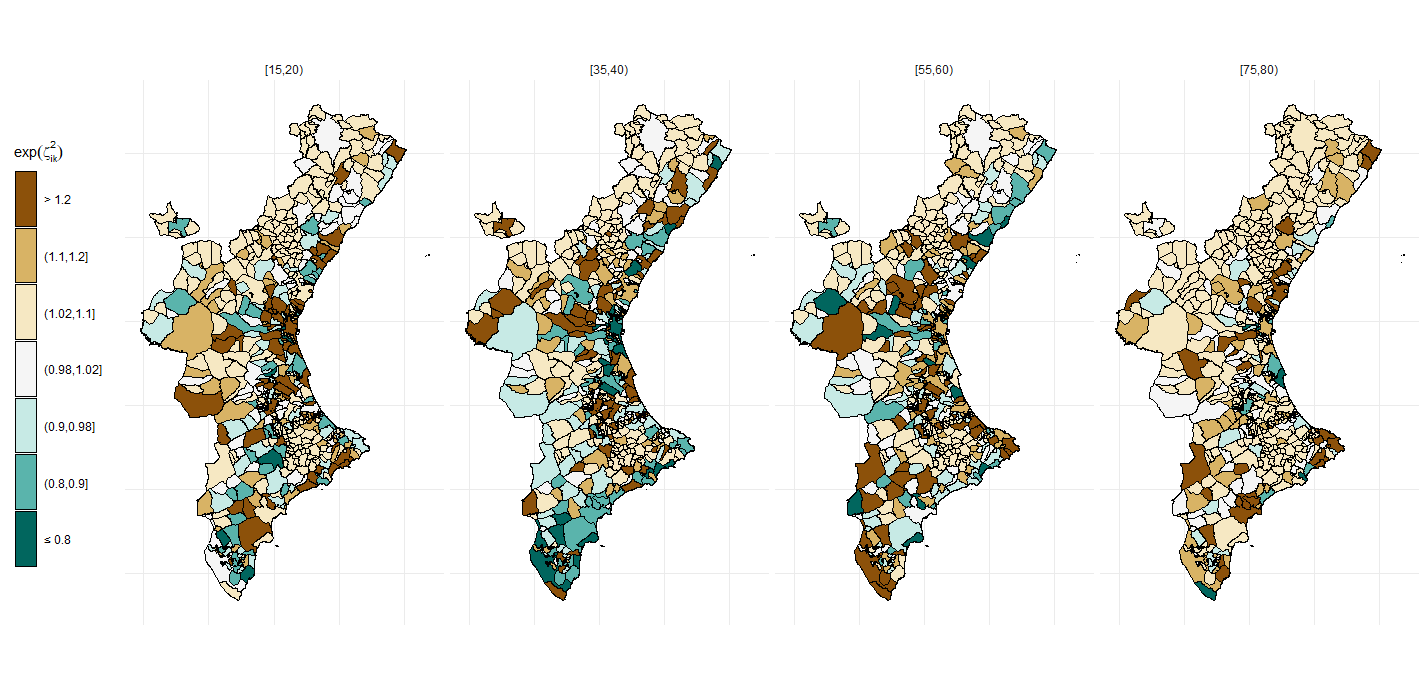}
    \caption{Posterior mean of the spatial-age interaction effect ($\exp(\bm{\zeta}^2)$) across the Valencian Community for the age groups [15,20), [35,40), [55,60), and [75,80).}
    \label{Fig:Spatial-age_effect}
\end{figure}

Figure \ref{Fig:Spatial-age_effect} shows the space-age effect $\exp(\zeta^2_{ik})$ corresponding to the age groups [15, 20), [35, 40), [55, 60), and [75, 80) (see the results for the remaining groups in the Supplementary Material). This random effect has been defined with a Type I structure, that is, without spatial or age correlation. As a result, a heterogeneous spatial pattern is observed, similar to the spatio-temporal effect described above. Between the groups [15, 20) and [35, 40), a reduction of the effect is observed in inland areas of the southern region, as well as in some inland areas of the central region. In contrast, the northern region shows a widespread positive contribution in both age groups. For the [55, 60) group, the space-age effect presents a notably heterogeneous pattern throughout the study region, without a clear spatial structure. Finally, in the [75, 80) group, a generalized increase in the effect is observed, with positive contributions predominating in most of the territory.\\
\begin{figure}[h!]
    \centering
    \includegraphics[scale=0.45]{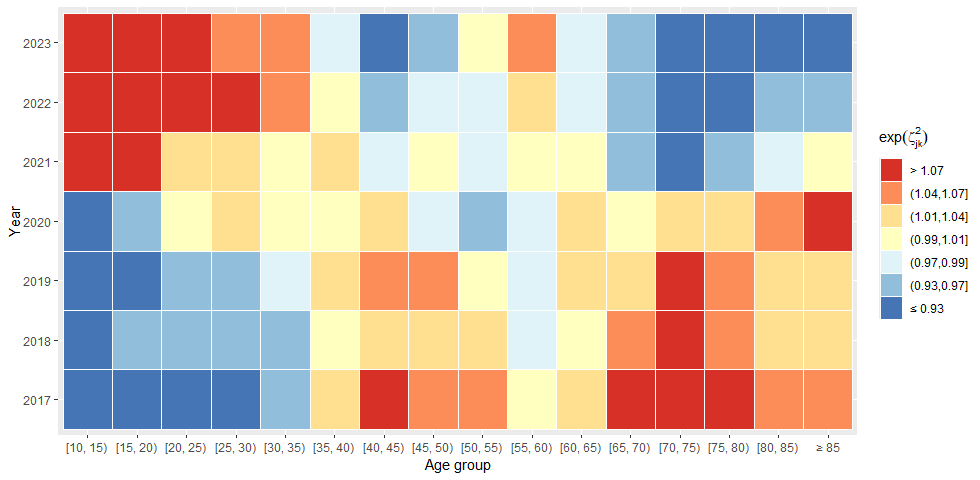}
    \caption{Posterior mean of the time–age interaction effect ($\exp(\bm{\zeta}^3)$) for the entire period 2017–2023 across all age groups, ranging from 15–19 years to 85 years and older.}
    \label{Fig:Time-age_effect}
\end{figure}

Finally, Figure \ref{Fig:Time-age_effect} shows the time-age effect $\exp(\zeta^3_{jk})$. This random effect has been modeled with a Type IV structure, which implies the presence of both temporal and age correlation. As a result, a smoothed pattern is observed in both dimensions: time and age. One region with a progressive increase in the rate of suicide-related emergency calls in the younger groups ([10, 35)) throughout the analyzed period is particularly notable. In these groups, a contribution to the reduction in call rates is observed between 2017 and 2020, followed by an increase in the effect (with values greater than unity) from 2021 onwards. This pattern intensifies in the youngest groups ([10,15) and [15,20)) and gradually decreases in the immediately higher groups. In contrast, the intermediate age groups ([35,55)) show the opposite behavior: they present positive contributions between 2017 and 2019, while from 2020 onwards, a progressive reduction in the effect is observed, with values less than 1. Finally, in the oldest age groups ([65,+)), the time-age effect is greater than 1 during the period 2017--2020, but decreases from 2021 onwards, indicating a possible reversal in the trend in suicide-related emergency call rates at older ages.\\

\section{Conclusion}
\label{Conclusion}

Indirect standardization remains a widely used technique in disease mapping, but its reliance on restrictive mathematical assumptions often leads to unreliable summary estimates. When the phenomenon under study (e.g., suicide risk) affects age strata heterogeneously, this method masks critical variations and interactions between age, space, and time. Furthermore, advances in statistical software, especially those based on INLA, have substantially improved the computational times required to estimate these types of models, making them a fast, easy-to-implement, and accessible option for many researchers.\\

Our analysis of suicide-related emergency call data demonstrates how conventional indirect standardization - when applied to phenomena with strong age-space-time interactions - can produce methodologically unsound results. The case study reveals that this approach generates biased risk estimates and unreliable conclusions when its underlying proportionality assumptions are violated, as occurs with suicide risk patterns. In contrast, we propose the use of age-structured hierarchical Bayesian models as a methodologically sound alternative, particularly relevant in contexts where data are disaggregated by multiple dimensions. This approach not only overcomes the limitations imposed by the proportionality assumption but also allows an estimate of individual spatial, temporal, and age effects, as well as their potential interactions. The decomposition obtained contributes to a more detailed understanding of the factors influencing the distribution of suicide-related emergency calls, allowing us to identify relevant patterns and potential hotspots, with important implications for the development of intervention and prevention strategies.\\

Suicide represents a multifactorial phenomenon whose incidence patterns, as reflected in mortality and emergency call data, exhibit significant spatial, temporal and age-specific variations. In the case of the Valencian Community, the results obtained reveal the existence of heterogeneous spatial patterns and interactions between space, time, and age, which could hardly have been identified using conventional methods applying indirect age standardization. Using this methodology, we observed an increase in the time effect in younger age groups starting in 2020, coinciding with the period following the COVID-19 pandemic, as well as spatial differences that suggest an unequal concentration of suicide-related emergency calls. These findings could point to the existence of specific dynamics that interact with age, region, and time, underscoring the need to adopt statistical approaches capable of capturing this complexity. In this sense, the proposed methodology not only allows for a more precise risk assessment but also represents a useful tool to guide public policies and more targeted and effective prevention strategies.\\

Although valuable for epidemiological research, our methodology shows particular promise for social science applications employing spatiotemporal analyses, especially where age effects demonstrate nonlinear and context-dependent patterns. This framework could provide critical insights into phenomena like substance abuse dynamics or gender-based violence, revealing how risk factors vary across age strata and interact with spatial and temporal determinants.

\section*{Acknowledgments}
This study was funded by the Social Observatory La Caixa Foundation (LCF/PR/SR21/52560010) and by the project PID2022-136455NB-I00, funded by Ministerio de Ciencia, Innovación y Universidades of Spain (MCIN/AEI/10.13039/501100011033/FEDER, UE) and the European Regional Development Fund.

\bibliographystyle{chicago}

\newpage
\section*{Sumplementary Material}\label{Suplementary_material}

\begin{table}[ht!]
\caption{Space-time-age models.}
\centering
\begin{tabular}{rlllllr}
  \hline
 Model & $\bm{\delta}$ & $\bm{\gamma}$ & $\bm{\zeta}^{1}$ & $\bm{\zeta}^{2}$ & $\bm{\zeta}^{3}$ & WAIC \\ 
  \hline
  1 & iid & - & - & - & - & 78841.19 \\ 
  2 & RW1 & - & - & - & - & 78840.97 \\ 
  3 & - & iid & - & - & - & 74362.06 \\ 
  4 & - & RW1 & - & - & - & 74362.26 \\ 
  5 & iid & iid & - & - & - & 70532.14 \\ 
  6 & RW1 & iid & - & - & - & 70531.87 \\ 
  7 & iid & RW1 & - & - & - & 70532.10 \\ 
  8 & RW1 & RW1 & - & - & - & 70531.86 \\ 
  9 & iid & - & Type I & - & - & 78164.59 \\ 
  10 & iid & - & Type II & - & - & 77858.79 \\ 
  11 & iid & - & Type III & - & - & 78155.39 \\ 
  12 & iid & - & Type IV & - & - & 77857.22 \\ 
  13 & RW1 & - & Type I & - & - & 78163.80 \\ 
  14 & RW1 & - & Type II & - & - & 77857.51 \\ 
  15 & RW1 & - & Type III & - & - & 78154.86 \\ 
  16 & RW1 & - & Type IV & - & - & 77854.97 \\ 
  17 & iid & iid & Type I & - & - & 69500.15 \\ 
  18 & RW1 & iid & Type I & - & - & 69498.66 \\ 
  19 & iid & RW1 & Type I & - & - & 69499.60 \\ 
  20 & RW1 & RW1 & Type I & - & - & 69500.31 \\ 
  21 & iid & iid & Type II & - & - & 69226.91 \\ 
  22 & RW1 & iid & Type II & - & - & 69225.91 \\ 
  23 & iid & RW1 & Type II & - & - & 69227.62 \\ 
  24 & RW1 & RW1 & Type II & - & - & 69227.54 \\ 
  25 & iid & iid & Type III & - & - & 69515.21 \\ 
  26 & RW1 & iid & Type III & - & - & 69514.44 \\ 
  27 & iid & RW1 & Type III & - & - & 69514.44 \\ 
  28 & RW1 & RW1 & Type III & - & - & 69514.41 \\ 
  29 & iid & iid & Type IV & - & - & 69257.86 \\ 
  30 & RW1 & iid & Type IV & - & - & 69256.25 \\ 
  31 & iid & RW1 & Type IV & - & - & 69257.71 \\ 
  32 & RW1 & RW1 & Type IV & - & - & 69256.22 \\ 
  33 & - & iid & - & Type I & - & 73563.72 \\ 
  34 & - & iid & - & Type II & - & 73412.09 \\ 
  35 & - & iid & - & Type III & - & 73559.02 \\ 
  36 & - & iid & - & Type IV & - & 73442.11 \\ 
  37 & - & RW1 & - & Type I & - & 73562.69 \\ 
  38 & - & RW1 & - & Type II & - & 73411.06 \\ 
  39 & - & RW1 & - & Type III & - & 73556.68 \\ 
  40 & - & RW1 & - & Type IV & - & 73441.59 \\ 
  41 & iid & iid & - & Type I & - & 69316.54 \\ 
  42 & RW1 & iid & - & Type I & - & 69315.63 \\ 
  43 & iid & RW1 & - & Type I & - & 69315.61 \\ 
  44 & RW1 & RW1 & - & Type I & - & 69314.43 \\ 
  45 & iid & iid & - & Type II & - & 69222.50 \\ 
  46 & RW1 & iid & - & Type II & - & 69223.06 \\
  47 & iid & RW1 & - & Type II & - & 69221.50 \\ 
  48 & RW1 & RW1 & - & Type II & - & 69220.76 \\ 
  49 & iid & iid & - & Type III & - & 69353.79 \\
  50 & RW1 & iid & - & Type III & - & 69354.39 \\ 
  \hline
\end{tabular}
\end{table}

\clearpage

\begin{table}[ht!]

\centering
\begin{tabular}{rlllllr}
\hline
 Model & $\bm{\delta}$ & $\bm{\gamma}$ & $\bm{\zeta}^{1}$ & $\bm{\zeta}^{2}$ & $\bm{\zeta}^{3}$ & WAIC \\ 
  \hline
  51 & iid & RW1 & - & Type III & - & 69353.38 \\ 
  52 & RW1 & RW1 & - & Type III & - & 69352.62 \\ 
  53 & iid & iid & - & Type IV & - & 69296.67 \\ 
  54 & RW1 & iid & - & Type IV & - & 69296.09 \\ 
  55 & iid & RW1 & - & Type IV & - & 69296.42 \\ 
  56 & RW1 & RW1 & - & Type IV & - & 69294.61 \\ 
  57 & iid & iid & - & - & Type I & 70447.70 \\ 
  58 & RW1 & iid & - & - & Type I & 70446.88 \\ 
  59 & iid & RW1 & - & - & Type I & 70447.81 \\ 
  60 & RW1 & RW1 & - & - & Type I & 70447.87 \\ 
  61 & iid & iid & - & - & Type II & 70410.92 \\ 
  62 & RW1 & iid & - & - & Type II & 70410.95 \\ 
  63 & iid & RW1 & - & - & Type II & 70411.08 \\ 
  64 & RW1 & RW1 & - & - & Type II & 70410.42 \\ 
  65 & iid & iid & - & - & Type III & 70410.35 \\ 
  66 & RW1 & iid & - & - & Type III & 70409.85 \\ 
  67 & iid & RW1 & - & - & Type III & 70411.29 \\ 
  68 & RW1 & RW1 & - & - & Type III & 70410.37 \\ 
  69 & iid & iid & - & - & Type IV & 70394.97 \\ 
  70 & RW1 & iid & - & - & Type IV & 70394.98 \\ 
  71 & iid & RW1 & - & - & Type IV & 70395.51 \\ 
  72 & RW1 & RW1 & - & - & Type IV & 70395.02 \\ 
  73 & iid & iid & Type I & Type I & - & 67974.39 \\ 
  74 & RW1 & iid & Type I & Type I & - & 67972.82 \\ 
  75 & iid & RW1 & Type I & Type I & - & 67971.99 \\ 
  76 & RW1 & RW1 & Type I & Type I & - & 67971.39 \\ 
  77 & iid & iid & Type II & Type I & - & 67718.39 \\ 
  78 & RW1 & iid & Type II & Type I & - & 67716.85 \\ 
  79 & iid & RW1 & Type II & Type I & - & 67716.76 \\ 
  80 & RW1 & RW1 & Type II & Type I & - & 67717.72 \\ 
  81 & iid & iid & Type I & Type II & - & 67942.45 \\ 
  82 & RW1 & iid & Type I & Type II & - & 67944.56 \\ 
  83 & iid & RW1 & Type I & Type II & - & 67945.63 \\ 
  84 & RW1 & RW1 & Type I & Type II & - & 67942.90 \\ 
  85 & iid & iid & Type II & Type II & - & 67684.72 \\ 
  86 & RW1 & iid & Type II & Type II & - & 67682.39 \\ 
  87 & iid & RW1 & Type II & Type II & - & 67683.04 \\ 
  88 & RW1 & RW1 & Type II & Type II & - & 67680.07 \\ 
  89 & iid & iid & Type III & Type I & - & 68016.94 \\ 
  90 & RW1 & iid & Type III & Type I & - & 68016.78 \\ 
  91 & iid & RW1 & Type III & Type I & - & 68015.22 \\ 
  92 & RW1 & RW1 & Type III & Type I & - & 68010.43 \\ 
  93 & iid & iid & Type III & Type II & - & 67983.30 \\ 
  94 & RW1 & iid & Type III & Type II & - & 67984.11 \\ 
  95 & iid & RW1 & Type III & Type II & - & 67981.96 \\ 
  96 & RW1 & RW1 & Type III & Type II & - & 67982.98 \\ 
  97 & iid & iid & Type I & Type III & - & 68024.43 \\ 
  98 & RW1 & iid & Type I & Type III & - & 68028.98 \\ 
  99 & iid & RW1 & Type I & Type III & - & 68025.37 \\ 
  100 & RW1 & RW1 & Type I & Type III & - & 68027.95 \\ 
  \hline
\end{tabular}
\end{table}

\clearpage

\begin{table}[ht!]

\centering
\begin{tabular}{rlllllr}
\hline
 Model & $\bm{\delta}$ & $\bm{\gamma}$ & $\bm{\zeta}^{1}$ & $\bm{\zeta}^{2}$ & $\bm{\zeta}^{3}$ & WAIC \\ 
  \hline
  101 & iid & iid & Type II & Type III & - & 67769.72 \\ 
  102 & RW1 & iid & Type II & Type III & - & 67762.54 \\ 
  103 & iid & RW1 & Type II & Type III & - & 67766.34 \\ 
  104 & RW1 & RW1 & Type II & Type III & - & 67767.02 \\ 
  105 & iid & iid & Type III & Type III & - & 68070.04 \\ 
  106 & RW1 & iid & Type III & Type III & - & 68069.52 \\ 
  107 & iid & RW1 & Type III & Type III & - & 68063.53 \\ 
  108 & RW1 & RW1 & Type III & Type III & - & 68064.79 \\ 
  109 & iid & iid & Type IV & Type I & - & 67769.42 \\ 
  110 & RW1 & iid & Type IV & Type I & - & 67768.32 \\ 
  111 & iid & RW1 & Type IV & Type I & - & 67767.05 \\ 
  112 & RW1 & RW1 & Type IV & Type I & - & 67768.63 \\ 
  113 & iid & iid & Type IV & Type II & - & 67768.63 \\ 
  114 & RW1 & iid & Type IV & Type II & - & 67732.77 \\ 
  115 & iid & RW1 & Type IV & Type II & - & 67731.89 \\ 
  116 & RW1 & RW1 & Type IV & Type II & - & 67732.25 \\ 
  117 & iid & iid & Type IV & Type III & - & 67819.17 \\ 
  118 & RW1 & iid & Type IV & Type III & - & 67819.26 \\ 
  119 & iid & RW1 & Type IV & Type III & - & 67818.81 \\ 
  120 & RW1 & RW1 & Type IV & Type III & - & 67819.64 \\ 
  121 & iid & iid & Type I & Type IV & - & 68031.21 \\ 
  122 & RW1 & iid & Type I & Type IV & - & 68027.45 \\ 
  123 & iid & RW1 & Type I & Type IV & - & 68027.12 \\ 
  124 & RW1 & RW1 & Type I & Type IV & - & 68028.87 \\ 
  125 & iid & iid & Type II & Type IV & - & 67771.18 \\ 
  126 & RW1 & iid & Type II & Type IV & - & 67769.38 \\ 
  127 & iid & RW1 & Type II & Type IV & - & 67768.65 \\ 
  128 & RW1 & RW1 & Type II & Type IV & - & 67768.68 \\ 
  129 & iid & iid & Type III & Type IV & - & 68066.71 \\ 
  130 & RW1 & iid & Type III & Type IV & - & 68066.27 \\ 
  131 & iid & RW1 & Type III & Type IV & - & 68062.18 \\ 
  132 & RW1 & RW1 & Type III & Type IV & - & 68063.68 \\ 
  133 & iid & iid & Type IV & Type IV & - & 67819.58 \\ 
  134 & RW1 & iid & Type IV & Type IV & - & 67816.63 \\ 
  135 & iid & RW1 & Type IV & Type IV & - & 67816.17 \\ 
  136 & RW1 & RW1 & Type IV & Type IV & - & 67815.21 \\ 
  137 & iid & iid & Type I & - & Type I & 69394.85 \\ 
  138 & RW1 & iid & Type I & - & Type I & 69394.53 \\ 
  139 & iid & RW1 & Type I & - & Type I & 69394.74 \\ 
  140 & RW1 & RW1 & Type I & - & Type I & 69394.37 \\ 
  141 & iid & iid & Type II & - & Type I & 69121.75 \\ 
  142 & RW1 & iid & Type II & - & Type I & 69121.57 \\ 
  143 & iid & RW1 & Type II & - & Type I & 69122.87 \\ 
  144 & RW1 & RW1 & Type II & - & Type I & 69122.10 \\ 
  145 & iid & iid & Type I & - & Type II & 69359.94 \\ 
  146 & RW1 & iid & Type I & - & Type II & 69360.02 \\ 
  147 & iid & RW1 & Type I & - & Type II & 69359.84 \\ 
  148 & RW1 & RW1 & Type I & - & Type II & 69359.52 \\ 
  149 & iid & iid & Type II & - & Type II & 69087.87 \\ 
  150 & RW1 & iid & Type II & - & Type II & 69087.35 \\
  \hline
\end{tabular}
\end{table}

\clearpage

\begin{table}[ht!]

\centering
\begin{tabular}{rlllllr}
\hline
 Model & $\bm{\delta}$ & $\bm{\gamma}$ & $\bm{\zeta}^{1}$ & $\bm{\zeta}^{2}$ & $\bm{\zeta}^{3}$ & WAIC \\ 
  \hline
 
  151 & iid & RW1 & Type II & - & Type II & 69087.90 \\ 
  152 & RW1 & RW1 & Type II & - & Type II & 69086.96 \\ 
  153 & iid & iid & Type III & - & Type I & 69411.31 \\ 
  154 & RW1 & iid & Type III & - & Type I & 69410.38 \\ 
  155 & iid & RW1 & Type III & - & Type I & 69411.69 \\ 
  156 & RW1 & RW1 & Type III & - & Type I & 69410.58 \\ 
  157 & iid & iid & Type III & - & Type II & 69376.53 \\ 
  158 & RW1 & iid & Type III & - & Type II & 69375.60 \\ 
  159 & iid & RW1 & Type III & - & Type II & 69376.29 \\ 
  160 & RW1 & RW1 & Type III & - & Type II & 69375.22 \\ 
  161 & iid & iid & Type I & - & Type III & 69358.87 \\
  162 & RW1 & iid & Type I & - & Type III & 69358.68 \\ 
  163 & iid & RW1 & Type I & - & Type III & 69359.62 \\ 
  164 & RW1 & RW1 & Type I & - & Type III & 69358.75 \\ 
  165 & iid & iid & Type II & - & Type III & 69087.27 \\ 
  166 & RW1 & iid & Type II & - & Type III & 69086.36 \\ 
  167 & iid & RW1 & Type II & - & Type III & 69087.53 \\ 
  168 & RW1 & RW1 & Type II & - & Type III & 69087.16 \\ 
  169 & iid & iid & Type III & - & Type III & 69375.11 \\ 
  170 & RW1 & iid & Type III & - & Type III & 69374.75 \\ 
  171 & iid & RW1 & Type III & - & Type III & 69375.86 \\ 
  172 & RW1 & RW1 & Type III & - & Type III & 69374.70 \\ 
  173 & iid & iid & Type IV & - & Type I & 69152.99 \\ 
  174 & RW1 & iid & Type IV & - & Type I & 69151.79 \\ 
  175 & iid & RW1 & Type IV & - & Type I & 69153.27 \\ 
  176 & RW1 & RW1 & Type IV & - & Type I & 69152.78 \\ 
  177 & iid & iid & Type IV & - & Type II & 69118.10 \\ 
  178 & RW1 & iid & Type IV & - & Type II & 69117.42 \\ 
  179 & iid & RW1 & Type IV & - & Type II & 69118.15 \\ 
  180 & RW1 & RW1 & Type IV & - & Type II & 69117.97 \\ 
  181 & iid & iid & Type IV & - & Type III & 69118.06 \\ 
  182 & RW1 & iid & Type IV & - & Type III & 69116.44 \\ 
  183 & iid & RW1 & Type IV & - & Type III & 69117.71 \\ 
  184 & RW1 & RW1 & Type IV & - & Type III & 69116.59 \\ 
  185 & iid & iid & Type I & - & Type IV & 69346.93 \\ 
  186 & RW1 & iid & Type I & - & Type IV & 69345.53 \\ 
  187 & iid & RW1 & Type I & - & Type IV & 69346.24 \\ 
  188 & RW1 & RW1 & Type I & - & Type IV & 69347.61 \\ 
  189 & iid & iid & Type II & - & Type IV & 69073.75 \\ 
  190 & RW1 & iid & Type II & - & Type IV & 69073.90 \\ 
  191 & iid & RW1 & Type II & - & Type IV & 69075.27 \\ 
  192 & RW1 & RW1 & Type II & - & Type IV & 69073.49 \\ 
  193 & iid & iid & Type III & - & Type IV & 69362.83 \\ 
  194 & RW1 & iid & Type III & - & Type IV & 69362.00 \\ 
  195 & iid & RW1 & Type III & - & Type IV & 69362.00 \\ 
  196 & RW1 & RW1 & Type III & - & Type IV & 69361.64 \\ 
  197 & iid & iid & Type IV & - & Type IV & 69103.96 \\ 
  198 & RW1 & iid & Type IV & - & Type IV & 69103.88 \\ 
  199 & iid & RW1 & Type IV & - & Type IV & 69104.77 \\ 
  200 & RW1 & RW1 & Type IV & - & Type IV & 69103.90 \\
  \hline
\end{tabular}
\end{table}

\clearpage

\begin{table}[ht!]

\centering
\begin{tabular}{rlllllr}
\hline
 Model & $\bm{\delta}$ & $\bm{\gamma}$ & $\bm{\zeta}^{1}$ & $\bm{\zeta}^{2}$ & $\bm{\zeta}^{3}$ & WAIC \\ 
  \hline
  201 & iid & iid & - & Type I & Type I & 69163.34 \\
  202 & RW1 & iid & - & Type I & Type I & 69162.37 \\ 
  203 & iid & RW1 & - & Type I & Type I & 69162.26 \\ 
  204 & RW1 & RW1 & - & Type I & Type I & 69161.96 \\ 
  205 & iid & iid & - & Type II & Type I & 69072.60 \\ 
  206 & RW1 & iid & - & Type II & Type I & 69071.14 \\ 
  207 & iid & RW1 & - & Type II & Type I & 69070.54 \\ 
  208 & RW1 & RW1 & - & Type II & Type I & 69070.40 \\ 
  209 & iid & iid & - & Type I & Type II & 69135.26 \\ 
  210 & RW1 & iid & - & Type I & Type II & 69136.14 \\ 
  211 & iid & RW1 & - & Type I & Type II & 69132.22 \\ 
  212 & RW1 & RW1 & - & Type I & Type II & 69134.14 \\ 
  213 & iid & iid & - & Type II & Type II & 69044.46 \\ 
  214 & RW1 & iid & - & Type II & Type II & 69043.01 \\ 
  215 & iid & RW1 & - & Type II & Type II & 69042.09 \\ 
  216 & RW1 & RW1 & - & Type II & Type II & 69042.41 \\ 
  217 & iid & iid & - & Type III & Type I & 69203.75 \\ 
  218 & RW1 & iid & - & Type III & Type I & 69202.75 \\ 
  219 & iid & RW1 & - & Type III & Type I & 69201.68 \\ 
  220 & RW1 & RW1 & - & Type III & Type I & 69203.37 \\ 
  221 & iid & iid & - & Type III & Type II & 69176.20 \\ 
  222 & RW1 & iid & - & Type III & Type II & 69175.12 \\ 
  223 & iid & RW1 & - & Type III & Type II & 69175.40 \\ 
  224 & RW1 & RW1 & - & Type III & Type II & 69175.48 \\ 
  225 & iid & iid & - & Type I & Type III & 69134.13 \\ 
  226 & RW1 & iid & - & Type I & Type III & 69135.02 \\ 
  227 & iid & RW1 & - & Type I & Type III & 69133.66 \\ 
  228 & RW1 & RW1 & - & Type I & Type III & 69133.29 \\ 
  229 & iid & iid & - & Type II & Type III & 69042.33 \\ 
  230 & RW1 & iid & - & Type II & Type III & 69042.12 \\ 
  231 & iid & RW1 & - & Type II & Type III & 69038.92 \\ 
  232 & RW1 & RW1 & - & Type II & Type III & 69040.94 \\ 
  233 & iid & iid & - & Type III & Type III & 69174.79 \\ 
  234 & RW1 & iid & - & Type III & Type III & 69175.42 \\ 
  235 & iid & RW1 & - & Type III & Type III & 69175.22 \\ 
  236 & RW1 & RW1 & - & Type III & Type III & 69172.64 \\ 
  237 & iid & iid & - & Type IV & Type I & 69153.28 \\ 
  238 & RW1 & iid & - & Type IV & Type I & 69152.27 \\ 
  239 & iid & RW1 & - & Type IV & Type I & 69150.11 \\ 
  240 & RW1 & RW1 & - & Type IV & Type I & 69152.00 \\ 
  241 & iid & iid & - & Type IV & Type II & 69124.84 \\
  242 & RW1 & iid & - & Type IV & Type II & 69125.81 \\ 
  243 & iid & RW1 & - & Type IV & Type II & 69121.40 \\ 
  244 & RW1 & RW1 & - & Type IV & Type II & 69123.55 \\ 
  245 & iid & iid & - & Type IV & Type III & 69124.30 \\ 
  246 & RW1 & iid & - & Type IV & Type III & 69122.27 \\ 
  247 & iid & RW1 & - & Type IV & Type III & 69122.98 \\ 
  248 & RW1 & RW1 & - & Type IV & Type III & 69122.06 \\ 
  249 & iid & iid & - & Type I & Type IV & 69128.37 \\ 
  250 & RW1 & iid & - & Type I & Type IV & 69126.62 \\ 
  \hline
\end{tabular}
\end{table}

\clearpage

\begin{table}[ht!]

\centering
\begin{tabular}{rlllllr}
\hline
 Model & $\bm{\delta}$ & $\bm{\gamma}$ & $\bm{\zeta}^{1}$ & $\bm{\zeta}^{2}$ & $\bm{\zeta}^{3}$ & WAIC \\ 
  \hline
   251 & iid & RW1 & - & Type I & Type IV & 69127.42 \\ 
  252 & RW1 & RW1 & - & Type I & Type IV & 69127.25 \\ 
  253 & iid & iid & - & Type II & Type IV & 69036.53 \\ 
  254 & RW1 & iid & - & Type II & Type IV & 69036.25 \\ 
  255 & iid & RW1 & - & Type II & Type IV & 69034.67 \\ 
  256 & RW1 & RW1 & - & Type II & Type IV & 69035.44 \\ 
  257 & iid & iid & - & Type III & Type IV & 69168.76 \\ 
  258 & RW1 & iid & - & Type III & Type IV & 69165.95 \\ 
  259 & iid & RW1 & - & Type III & Type IV & 69167.21 \\ 
  260 & RW1 & RW1 & - & Type III & Type IV & 69166.74 \\ 
  261 & iid & iid & - & Type IV & Type IV & 69116.73 \\ 
  262 & RW1 & iid & - & Type IV & Type IV & 69116.94 \\ 
  263 & iid & RW1 & - & Type IV & Type IV & 69115.63 \\ 
  264 & RW1 & RW1 & - & Type IV & Type IV & 69114.23 \\ 
  265 & iid & iid & Type I & Type I & Type I & 67793.40 \\ 
  266 & RW1 & iid & Type I & Type I & Type I & 67793.21 \\ 
  267 & iid & RW1 & Type I & Type I & Type I & 67793.51 \\ 
  268 & RW1 & RW1 & Type I & Type I & Type I & 67791.45 \\ 
  269 & iid & iid & Type II & Type I & Type I & 67536.66 \\ 
  270 & RW1 & iid & Type II & Type I & Type I & 67536.33 \\ 
  271 & iid & RW1 & Type II & Type I & Type I & 67534.65 \\ 
  272 & RW1 & RW1 & Type II & Type I & Type I & 67534.81 \\ 
  273 & iid & iid & Type III & Type I & Type I & 67838.49 \\ 
  274 & RW1 & iid & Type III & Type I & Type I & 67834.60 \\ 
  275 & iid & RW1 & Type III & Type I & Type I & 67835.38 \\ 
  276 & RW1 & RW1 & Type III & Type I & Type I & 67836.64 \\ 
  277 & iid & iid & Type IV & Type I & Type I & 67588.35 \\ 
  278 & RW1 & iid & Type IV & Type I & Type I & 67588.74 \\ 
  279 & iid & RW1 & Type IV & Type I & Type I & 67589.23 \\ 
  280 & RW1 & RW1 & Type IV & Type I & Type I & 67587.87 \\ 
  281 & iid & iid & Type I & Type II & Type I & 67772.15 \\
  282 & RW1 & iid & Type I & Type II & Type I & 67770.88 \\ 
  283 & iid & RW1 & Type I & Type II & Type I & 67770.80 \\ 
  284 & RW1 & RW1 & Type I & Type II & Type I & 67770.73 \\ 
  285 & iid & iid & Type I & Type III & Type I & 67848.14 \\ 
  286 & RW1 & iid & Type I & Type III & Type I & 67847.69 \\ 
  287 & iid & RW1 & Type I & Type III & Type I & 67846.01 \\ 
  288 & RW1 & RW1 & Type I & Type III & Type I & 67842.39 \\ 
  289 & iid & iid & Type I & Type IV & Type I & 67855.37 \\ 
  290 & RW1 & iid & Type I & Type IV & Type I & 67855.62 \\ 
  291 & iid & RW1 & Type I & Type IV & Type I & 67853.60 \\ 
  292 & RW1 & RW1 & Type I & Type IV & Type I & 67856.16 \\ 
  293 & iid & iid & Type I & Type I & Type II & 67768.68 \\ 
  294 & RW1 & iid & Type I & Type I & Type II & 67767.92 \\ 
  295 & iid & RW1 & Type I & Type I & Type II & 67766.98 \\ 
  296 & RW1 & RW1 & Type I & Type I & Type II & 67767.31 \\ 
  297 & iid & iid & Type I & Type I & Type III & 67766.82 \\ 
  298 & RW1 & iid & Type I & Type I & Type III & 67767.17 \\ 
  299 & iid & RW1 & Type I & Type I & Type III & 67765.79 \\ 
  300 & RW1 & RW1 & Type I & Type I & Type III & 67764.68 \\
\hline
\end{tabular}
\end{table}

\clearpage

\begin{table}[ht!]

\centering
\begin{tabular}{rlllllr}
\hline
 Model & $\bm{\delta}$ & $\bm{\gamma}$ & $\bm{\zeta}^{1}$ & $\bm{\zeta}^{2}$ & $\bm{\zeta}^{3}$ & WAIC \\ 
  \hline
  301 & iid & iid & Type I & Type I & Type IV & 67764.69 \\ 
  302 & RW1 & iid & Type I & Type I & Type IV & 67762.34 \\ 
  303 & iid & RW1 & Type I & Type I & Type IV & 67764.32 \\ 
  304 & RW1 & RW1 & Type I & Type I & Type IV & 67763.71 \\ 
  305 & iid & iid & Type II & Type II & Type I & 67507.16 \\ 
  306 & RW1 & iid & Type II & Type II & Type I & 67509.19 \\ 
  307 & iid & RW1 & Type II & Type II & Type I & 67507.71 \\ 
  308 & RW1 & RW1 & Type II & Type II & Type I & 67508.17 \\ 
  309 & iid & iid & Type II & Type III & Type I & 67588.92 \\ 
  310 & RW1 & iid & Type II & Type III & Type I & 67588.20 \\ 
  311 & iid & RW1 & Type II & Type III & Type I & 67588.56 \\ 
  312 & RW1 & RW1 & Type II & Type III & Type I & 67585.42 \\ 
  313 & iid & iid & Type II & Type IV & Type I & 67595.05 \\ 
  314 & RW1 & iid & Type II & Type IV & Type I & 67594.65 \\ 
  315 & iid & RW1 & Type II & Type IV & Type I & 67593.83 \\ 
  316 & RW1 & RW1 & Type II & Type IV & Type I & 67594.91 \\ 
  317 & iid & iid & Type II & Type I & Type II & 67510.86 \\ 
  318 & RW1 & iid & Type II & Type I & Type II & 67511.67 \\ 
  319 & iid & RW1 & Type II & Type I & Type II & 67510.80 \\ 
  320 & RW1 & RW1 & Type II & Type I & Type II & 67510.00 \\ 
  321 & iid & iid & Type II & Type I & Type III & 67510.26 \\
  322 & RW1 & iid & Type II & Type I & Type III & 67510.26 \\ 
  323 & iid & RW1 & Type II & Type I & Type III & 67508.37 \\ 
  324 & RW1 & RW1 & Type II & Type I & Type III & 67508.62 \\ 
  325 & iid & iid & Type II & Type I & Type IV & 67507.71 \\ 
  326 & RW1 & iid & Type II & Type I & Type IV & 67507.88 \\ 
  327 & iid & RW1 & Type II & Type I & Type IV & 67505.20 \\ 
  328 & RW1 & RW1 & Type II & Type I & Type IV & 67506.02 \\ 
  329 & iid & iid & Type III & Type II & Type I & 67809.50 \\ 
  330 & RW1 & iid & Type III & Type II & Type I & 67807.76 \\ 
  331 & iid & RW1 & Type III & Type II & Type I & 67804.96 \\ 
  332 & RW1 & RW1 & Type III & Type II & Type I & 67810.80 \\ 
  333 & iid & iid & Type III & Type III & Type I & 67892.40 \\ 
  334 & RW1 & iid & Type III & Type III & Type I & 67890.55 \\ 
  335 & iid & RW1 & Type III & Type III & Type I & 67891.69 \\ 
  336 & RW1 & RW1 & Type III & Type III & Type I & 67890.19 \\ 
  337 & iid & iid & Type III & Type IV & Type I & 67894.41 \\ 
  338 & RW1 & iid & Type III & Type IV & Type I & 67886.75 \\ 
  339 & iid & RW1 & Type III & Type IV & Type I & 67893.97 \\ 
  340 & RW1 & RW1 & Type III & Type IV & Type I & 67893.79 \\ 
  341 & iid & iid & Type III & Type I & Type II & 67813.13 \\ 
  342 & RW1 & iid & Type III & Type I & Type II & 67811.88 \\ 
  343 & iid & RW1 & Type III & Type I & Type II & 67811.90 \\ 
  344 & RW1 & RW1 & Type III & Type I & Type II & 67806.89 \\ 
  345 & iid & iid & Type III & Type I & Type III & 67810.85 \\ 
  346 & RW1 & iid & Type III & Type I & Type III & 67812.58 \\ 
  347 & iid & RW1 & Type III & Type I & Type III & 67809.74 \\ 
  348 & RW1 & RW1 & Type III & Type I & Type III & 67809.02 \\ 
  349 & iid & iid & Type III & Type I & Type IV & 67808.76 \\ 
  350 & RW1 & iid & Type III & Type I & Type IV & 67807.76 \\
  \hline
\end{tabular}
\end{table}

\clearpage

\begin{table}[ht!]

\centering
\begin{tabular}{rlllllr}
\hline
 Model & $\bm{\delta}$ & $\bm{\gamma}$ & $\bm{\zeta}^{1}$ & $\bm{\zeta}^{2}$ & $\bm{\zeta}^{3}$ & WAIC \\ 
  \hline
  351 & iid & RW1 & Type III & Type I & Type IV & 67808.63 \\ 
  352 & RW1 & RW1 & Type III & Type I & Type IV & 67807.09 \\ 
  353 & iid & iid & Type IV & Type II & Type I & 67559.02 \\ 
  354 & RW1 & iid & Type IV & Type II & Type I & 67557.89 \\ 
  355 & iid & RW1 & Type IV & Type II & Type I & 67557.16 \\ 
  356 & RW1 & RW1 & Type IV & Type II & Type I & 67557.49 \\ 
  357 & iid & iid & Type IV & Type III & Type I & 67642.11 \\ 
  358 & RW1 & iid & Type IV & Type III & Type I & 67641.34 \\ 
  359 & iid & RW1 & Type IV & Type III & Type I & 67641.35 \\ 
  360 & RW1 & RW1 & Type IV & Type III & Type I & 67641.23 \\ 
  361 & iid & iid & Type IV & Type IV & Type I & 67645.50 \\
  362 & RW1 & iid & Type IV & Type IV & Type I & 67645.67 \\ 
  363 & iid & RW1 & Type IV & Type IV & Type I & 67885.24 \\ 
  364 & RW1 & RW1 & Type IV & Type IV & Type I & 67885.07 \\ 
  365 & iid & iid & Type IV & Type I & Type II & 67796.44 \\ 
  366 & RW1 & iid & Type IV & Type I & Type II & 67799.38 \\ 
  367 & iid & RW1 & Type IV & Type I & Type II & 67798.43 \\ 
  368 & RW1 & RW1 & Type IV & Type I & Type II & 67796.07 \\ 
  369 & iid & iid & Type IV & Type I & Type III & 67796.66 \\ 
  370 & RW1 & iid & Type IV & Type I & Type III & 67799.52 \\ 
  371 & iid & RW1 & Type IV & Type I & Type III & 67794.67 \\ 
  372 & RW1 & RW1 & Type IV & Type I & Type III & 67794.85 \\ 
  373 & iid & iid & Type IV & Type I & Type IV & 67557.09 \\ 
  374 & RW1 & iid & Type IV & Type I & Type IV & 67804.69 \\ 
  375 & iid & RW1 & Type IV & Type I & Type IV & 67803.00 \\ 
  376 & RW1 & RW1 & Type IV & Type I & Type IV & 67798.10 \\ 
  377 & iid & iid & Type I & Type II & Type II & 67978.47 \\ 
  378 & RW1 & iid & Type I & Type II & Type II & 67979.94 \\ 
  379 & iid & RW1 & Type I & Type II & Type II & 67976.78 \\ 
  380 & RW1 & RW1 & Type I & Type II & Type II & 67972.94 \\ 
  381 & iid & iid & Type I & Type III & Type II & 68099.67 \\ 
  382 & RW1 & iid & Type I & Type III & Type II & 68098.98 \\ 
  383 & iid & RW1 & Type I & Type III & Type II & 68096.70 \\ 
  384 & RW1 & RW1 & Type I & Type III & Type II & 68096.07 \\ 
  385 & iid & iid & Type I & Type IV & Type II & 68165.11 \\ 
  386 & RW1 & iid & Type I & Type IV & Type II & 68164.72 \\ 
  387 & iid & RW1 & Type I & Type IV & Type II & 68161.98 \\ 
  388 & RW1 & RW1 & Type I & Type IV & Type II & 68159.67 \\ 
  389 & iid & iid & Type II & Type II & Type II & 67767.55 \\ 
  390 & RW1 & iid & Type II & Type II & Type II & 67768.50 \\ 
  391 & iid & RW1 & Type II & Type II & Type II & 67764.43 \\ 
  392 & RW1 & RW1 & Type II & Type II & Type II & 67762.51 \\ 
  393 & iid & iid & Type II & Type III & Type II & 67812.40 \\ 
  394 & RW1 & iid & Type II & Type III & Type II & 67807.78 \\ 
  395 & iid & RW1 & Type II & Type III & Type II & 67804.46 \\ 
  396 & RW1 & RW1 & Type II & Type III & Type II & 67804.67 \\ 
  397 & iid & iid & Type II & Type IV & Type II & 67884.87 \\ 
  398 & RW1 & iid & Type II & Type IV & Type II & 67893.14 \\ 
  399 & iid & RW1 & Type II & Type IV & Type II & 67875.95 \\ 
  400 & RW1 & RW1 & Type II & Type IV & Type II & 67886.97 \\
    \hline
\end{tabular}
\end{table}

\clearpage

\begin{table}[ht!]

\centering
\begin{tabular}{rlllllr}
\hline
 Model & $\bm{\delta}$ & $\bm{\gamma}$ & $\bm{\zeta}^{1}$ & $\bm{\zeta}^{2}$ & $\bm{\zeta}^{3}$ & WAIC \\ 
  \hline
  401 & iid & iid & Type III & Type II & Type II & 68099.47 \\
  402 & RW1 & iid & Type III & Type II & Type II & 68099.33 \\ 
  403 & iid & RW1 & Type III & Type II & Type II & 68097.15 \\ 
  404 & RW1 & RW1 & Type III & Type II & Type II & 68096.77 \\ 
  405 & iid & iid & Type III & Type III & Type II & 68135.93 \\ 
  406 & RW1 & iid & Type III & Type III & Type II & 68136.96 \\ 
  407 & iid & RW1 & Type III & Type III & Type II & 68136.47 \\ 
  408 & RW1 & RW1 & Type III & Type III & Type II & 68133.23 \\ 
  409 & iid & iid & Type III & Type IV & Type II & 68097.96 \\ 
  410 & RW1 & iid & Type III & Type IV & Type II & 68098.37 \\ 
  411 & iid & RW1 & Type III & Type IV & Type II & 68097.28 \\ 
  412 & RW1 & RW1 & Type III & Type IV & Type II & 68094.38 \\ 
  413 & iid & iid & Type IV & Type II & Type II & 67853.29 \\ 
  414 & RW1 & iid & Type IV & Type II & Type II & 67853.44 \\ 
  415 & iid & RW1 & Type IV & Type II & Type II & 67851.54 \\ 
  416 & RW1 & RW1 & Type IV & Type II & Type II & 67852.79 \\ 
  417 & iid & iid & Type IV & Type III & Type II & 67823.92 \\ 
  418 & RW1 & iid & Type IV & Type III & Type II & 67822.13 \\ 
  419 & iid & RW1 & Type IV & Type III & Type II & 67820.79 \\ 
  420 & RW1 & RW1 & Type IV & Type III & Type II & 67830.94 \\ 
  421 & iid & iid & Type IV & Type IV & Type II & 67862.82 \\ 
  422 & RW1 & iid & Type IV & Type IV & Type II & 67863.51 \\ 
  423 & iid & RW1 & Type IV & Type IV & Type II & 67861.57 \\ 
  424 & RW1 & RW1 & Type IV & Type IV & Type II & 67861.08 \\ 
  425 & iid & iid & Type I & Type II & Type III & 67973.14 \\ 
  426 & RW1 & iid & Type I & Type II & Type III & 67972.33 \\ 
  427 & iid & RW1 & Type I & Type II & Type III & 67970.16 \\ 
  428 & RW1 & RW1 & Type I & Type II & Type III & 67969.60 \\ 
  429 & iid & iid & Type I & Type III & Type III & 68098.96 \\ 
  430 & RW1 & iid & Type I & Type III & Type III & 68098.36 \\ 
  431 & iid & RW1 & Type I & Type III & Type III & 68095.84 \\ 
  432 & RW1 & RW1 & Type I & Type III & Type III & 68095.99 \\ 
  433 & iid & iid & Type I & Type IV & Type III & 68159.30 \\ 
  434 & RW1 & iid & Type I & Type IV & Type III & 68159.70 \\ 
  435 & iid & RW1 & Type I & Type IV & Type III & 68155.97 \\ 
  436 & RW1 & RW1 & Type I & Type IV & Type III & 68155.61 \\ 
  437 & iid & iid & Type II & Type II & Type III & 67762.60 \\ 
  438 & RW1 & iid & Type II & Type II & Type III & 67762.50 \\ 
  439 & iid & RW1 & Type II & Type II & Type III & 67760.57 \\ 
  440 & RW1 & RW1 & Type II & Type II & Type III & 67761.95 \\ 
  441 & iid & iid & Type II & Type III & Type III & 67810.29 \\
  442 & RW1 & iid & Type II & Type III & Type III & 67808.38 \\ 
  443 & iid & RW1 & Type II & Type III & Type III & 67804.61 \\ 
  444 & RW1 & RW1 & Type II & Type III & Type III & 67806.68 \\ 
  445 & iid & iid & Type II & Type IV & Type III & 67880.14 \\ 
  446 & RW1 & iid & Type II & Type IV & Type III & 67860.70 \\ 
  447 & iid & RW1 & Type II & Type IV & Type III & 67877.51 \\ 
  448 & RW1 & RW1 & Type II & Type IV & Type III & 67849.63 \\ 
  449 & iid & iid & Type III & Type II & Type III & 68098.79 \\ 
  450 & RW1 & iid & Type III & Type II & Type III & 68097.78 \\
  \hline
\end{tabular}
\end{table}

\clearpage

\begin{table}[ht!]

\centering
\begin{tabular}{rlllllr}
\hline
 Model & $\bm{\delta}$ & $\bm{\gamma}$ & $\bm{\zeta}^{1}$ & $\bm{\zeta}^{2}$ & $\bm{\zeta}^{3}$ & WAIC \\ 
  \hline
   
  451 & iid & RW1 & Type III & Type II & Type III & 68096.87 \\ 
  452 & RW1 & RW1 & Type III & Type II & Type III & 68096.69 \\ 
  453 & iid & iid & Type III & Type III & Type III & 68133.86 \\ 
  454 & RW1 & iid & Type III & Type III & Type III & 68139.48 \\ 
  455 & iid & RW1 & Type III & Type III & Type III & 68133.24 \\ 
  456 & RW1 & RW1 & Type III & Type III & Type III & 68132.71 \\ 
  457 & iid & iid & Type III & Type IV & Type III & 68097.79 \\ 
  458 & RW1 & iid & Type III & Type IV & Type III & 68097.55 \\ 
  459 & iid & RW1 & Type III & Type IV & Type III & 68097.35 \\ 
  460 & RW1 & RW1 & Type III & Type IV & Type III & 68094.35 \\ 
  461 & iid & iid & Type IV & Type II & Type III & 67856.63 \\ 
  462 & RW1 & iid & Type IV & Type II & Type III & 67850.73 \\ 
  463 & iid & RW1 & Type IV & Type II & Type III & 67854.69 \\ 
  464 & RW1 & RW1 & Type IV & Type II & Type III & 67848.18 \\ 
  465 & iid & iid & Type IV & Type III & Type III & 67820.59 \\ 
  466 & RW1 & iid & Type IV & Type III & Type III & 67819.40 \\ 
  467 & iid & RW1 & Type IV & Type III & Type III & 67821.11 \\ 
  468 & RW1 & RW1 & Type IV & Type III & Type III & 67819.79 \\ 
  469 & iid & iid & Type IV & Type IV & Type III & 67862.04 \\ 
  470 & RW1 & iid & Type IV & Type IV & Type III & 67861.25 \\ 
  471 & iid & RW1 & Type IV & Type IV & Type III & 67859.49 \\ 
  472 & RW1 & RW1 & Type IV & Type IV & Type III & 67859.56 \\ 
  473 & iid & iid & Type I & Type II & Type IV & 67976.78 \\ 
  474 & RW1 & iid & Type I & Type II & Type IV & 67981.36 \\ 
  475 & iid & RW1 & Type I & Type II & Type IV & 67976.96 \\ 
  476 & RW1 & RW1 & Type I & Type II & Type IV & 67976.83 \\ 
  477 & iid & iid & Type I & Type III & Type IV & 68100.77 \\ 
  478 & RW1 & iid & Type I & Type III & Type IV & 68099.43 \\ 
  479 & iid & RW1 & Type I & Type III & Type IV & 68097.89 \\ 
  480 & RW1 & RW1 & Type I & Type III & Type IV & 68097.06 \\ 
  481 & iid & iid & Type I & Type IV & Type IV & 68168.38 \\
  482 & RW1 & iid & Type I & Type IV & Type IV & 68166.72 \\ 
  483 & iid & RW1 & Type I & Type IV & Type IV & 68162.30 \\ 
  484 & RW1 & RW1 & Type I & Type IV & Type IV & 68160.24 \\ 
  485 & iid & iid & Type II & Type II & Type IV & 67770.66 \\ 
  486 & RW1 & iid & Type II & Type II & Type IV & 67768.29 \\ 
  487 & iid & RW1 & Type II & Type II & Type IV & 67764.97 \\ 
  488 & RW1 & RW1 & Type II & Type II & Type IV & 67763.73 \\ 
  489 & iid & iid & Type II & Type III & Type IV & 67817.18 \\ 
  490 & RW1 & iid & Type II & Type III & Type IV & 67812.73 \\ 
  491 & iid & RW1 & Type II & Type III & Type IV & 67813.83 \\ 
  492 & RW1 & RW1 & Type II & Type III & Type IV & 67811.41 \\ 
  493 & iid & iid & Type II & Type IV & Type IV & 67856.51 \\ 
  494 & RW1 & iid & Type II & Type IV & Type IV & 67861.56 \\ 
  495 & iid & RW1 & Type II & Type IV & Type IV & 67853.21 \\ 
  496 & RW1 & RW1 & Type II & Type IV & Type IV & 67854.19 \\ 
  497 & iid & iid & Type III & Type II & Type IV & 68099.10 \\ 
  498 & RW1 & iid & Type III & Type II & Type IV & 68098.45 \\ 
  499 & iid & RW1 & Type III & Type II & Type IV & 68096.61 \\ 
  500 & RW1 & RW1 & Type III & Type II & Type IV & 68096.96 \\
  \hline
\end{tabular}
\end{table}

\clearpage

\begin{table}[ht!]

\centering
\begin{tabular}{rlllllr}
\hline
 Model & $\bm{\delta}$ & $\bm{\gamma}$ & $\bm{\zeta}^{1}$ & $\bm{\zeta}^{2}$ & $\bm{\zeta}^{3}$ & WAIC \\ 
  \hline
  501 & iid & iid & Type III & Type III & Type IV & 68139.67 \\ 
  502 & RW1 & iid & Type III & Type III & Type IV & 68130.72 \\ 
  503 & iid & RW1 & Type III & Type III & Type IV & 68141.46 \\ 
  504 & RW1 & RW1 & Type III & Type III & Type IV & 68138.22 \\ 
  505 & iid & iid & Type III & Type IV & Type IV & 68096.52 \\ 
  506 & RW1 & iid & Type III & Type IV & Type IV & 68095.53 \\ 
  507 & iid & RW1 & Type III & Type IV & Type IV & 68094.51 \\ 
  508 & RW1 & RW1 & Type III & Type IV & Type IV & 68091.35 \\ 
  509 & iid & iid & Type IV & Type II & Type IV & 67846.63 \\ 
  510 & RW1 & iid & Type IV & Type II & Type IV & 67846.73 \\ 
  511 & iid & RW1 & Type IV & Type II & Type IV & 67850.62 \\ 
  512 & RW1 & RW1 & Type IV & Type II & Type IV & 67845.12 \\ 
  513 & iid & iid & Type IV & Type III & Type IV & 67819.61 \\ 
  514 & RW1 & iid & Type IV & Type III & Type IV & 67815.23 \\ 
  515 & iid & RW1 & Type IV & Type III & Type IV & 67811.55 \\ 
  516 & RW1 & RW1 & Type IV & Type III & Type IV & 67812.10 \\ 
  517 & iid & iid & Type IV & Type IV & Type IV & 67859.04 \\ 
  518 & RW1 & iid & Type IV & Type IV & Type IV & 67858.35 \\ 
  519 & iid & RW1 & Type IV & Type IV & Type IV & 67856.55 \\ 
  520 & RW1 & RW1 & Type IV & Type IV & Type IV & 67856.38 \\ 
   \hline
\end{tabular}
\end{table}

\newpage
\begin{landscape}
\begin{figure}[ht]
    \centering
    \includegraphics[scale=0.7]{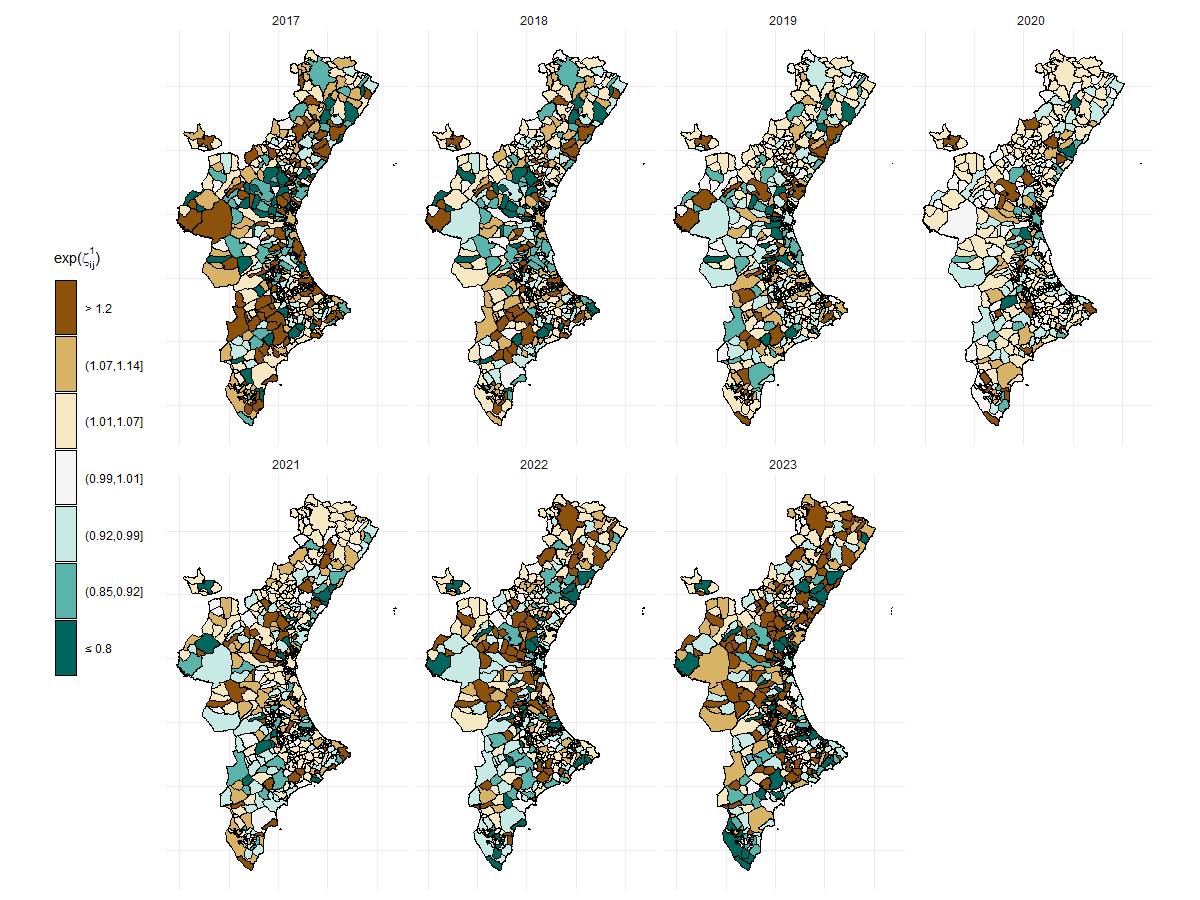}
    \caption{Posterior mean of the spatio-temporal effect ($\exp(\bm{\zeta}^1)$) across the Valencian Community for the complete period.}
    \label{Fig:Spatio-temporal_effect}
\end{figure}
\end{landscape}

\newpage
\begin{landscape}
\begin{figure}[ht]
    \centering
    \includegraphics[scale=0.6]    {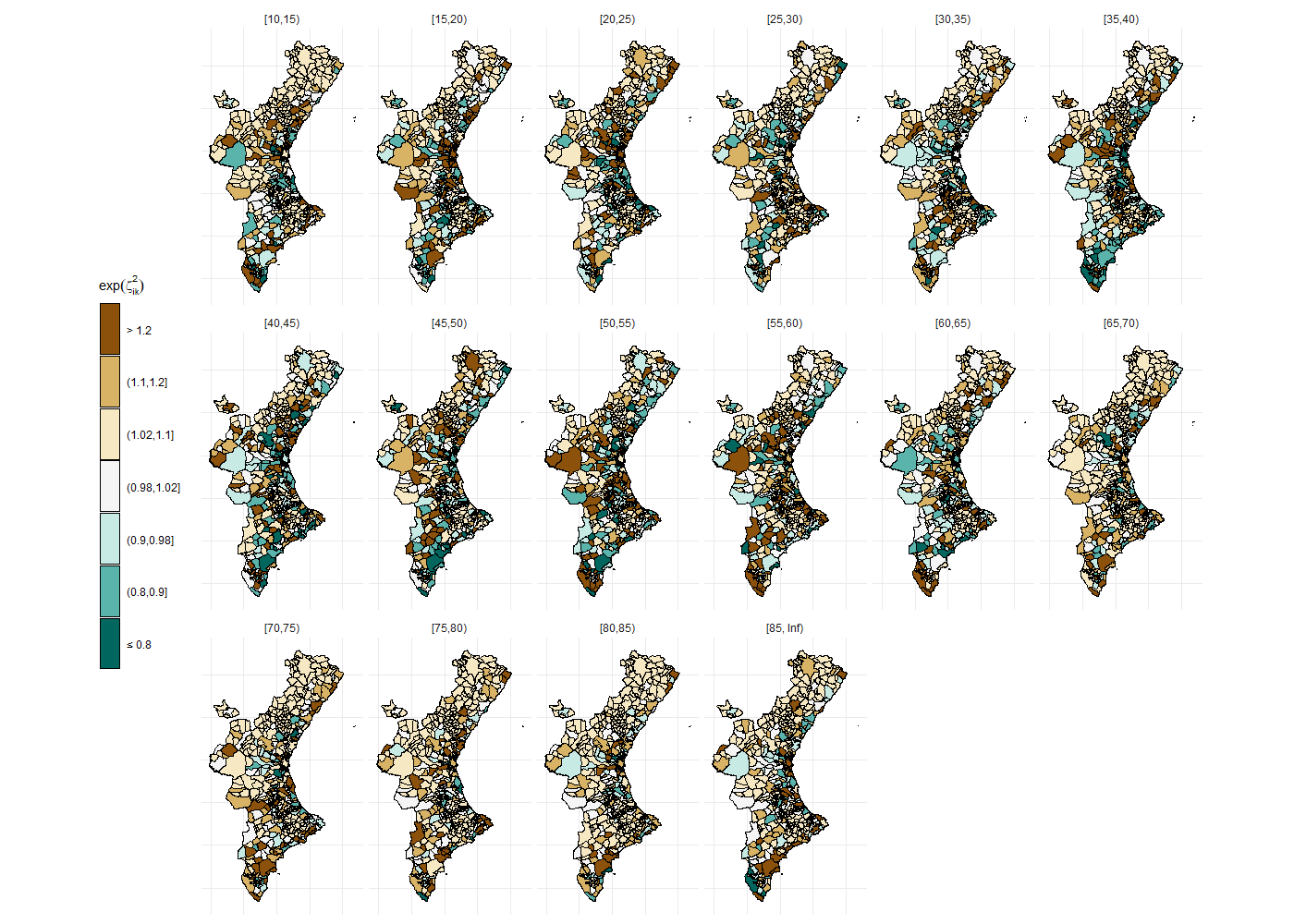}
    \caption{Posterior mean of the spatial-age effect ($\exp(\bm{\zeta}^2)$) across the Valencian Community for all age groups.}
    \label{Fig:Spatio-temporal_effect}
\end{figure}
\end{landscape}

\end{document}